\documentclass[journal,onecolumn,12pt]{IEEEtran}

\ifCLASSINFOpdf
\else
\fi
\pdfoutput=1
\usepackage[pdftex]{graphicx}
\usepackage{color}
\usepackage{bm}
\usepackage{subfigure}
\usepackage{cite}

\input alphabet
\input abrege
\def\cro#1{\left[#1\right]}                
               
\def\norm#1{\left\|#1\right\|}

\def\ST{\text{s.t.\:}}

\newsavebox{\fminibox}
\newlength{\fminilength}
\newenvironment{fminipage}[1][\linewidth]
  {\setlength{\fminilength}{#1}
   \begin{lrbox}{\fminibox}\begin{minipage}{\fminilength}}
  {\end{minipage}\end{lrbox}\noindent\fbox{\usebox{\fminibox}}}

	\def\pmu{^{-1}}

   \def\wh#1{\widehat{#1}}                 

	\def\T{^\tD} 

   \def\argmin{\mathop{\mathrm{arg\,min}}} 
   \def\rond#1{\overset{\kern-0.33em~_\circ}{#1}}
   \def\rondit[#1]#2{\overset{\kern#1~_\circ}{#2}}

\def\babs{\begin{abstract}}             \def\eabs{\end{abstract}}
\def\barr{\begin{array}}                \def\earr{\end{array}}
\def\bcc{\begin{center}}                \def\ecc{\end{center}}
\def\bdes{\begin{description}}          \def\edes{\end{description}}
\def\bdoc{\begin{document}}             \def\edoc{\end{document}}
\def\ben{\begin{enumerate}}             \def\een{\end{enumerate}}
\def\barr{\begin{array}}             \def\earr{\end{array}}
\def\beqn{\begin{eqnarray}}             \def\eeqn{\end{eqnarray}}
\def\beqnx{\begin{eqnarray*}}           \def\eeqnx{\end{eqnarray*}}
\def\bseqn{\begin{subeqnarray}}         \def\eseqn{\end{subeqnarray}}
\def\beq#1\eeq{\begin{equation}#1   \end{equation}}
\def\bal#1\eal{\begin{align}#1\end{align}}
\def\balx#1\ealx{\begin{align*}#1\end{align*}}
\def\beqx{$$}                           \def\eeqx{$$}
\def\bfig{\protect\begin{figure}}       \def\efig{\protect\end{figure}}
\def\bfigx{\protect\begin{figure*}}     \def\efigx{\protect\end{figure*}}
\def\bfigt{\protect\begin{figurette}}   \def\efigt{\protect\end{figurette}}
\def\bfl{\begin{flushleft}}             \def\efl{\end{flushleft}}
\def\bfr{\begin{flushright}}            \def\efr{\end{flushright}}
\def\bit{\begin{itemize}}               \def\eit{\end{itemize}}
\def\bmi{\begin{minipage}}              \def\emi{\end{minipage}}
\def\bfmi{\begin{fminipage}}            \def\efmi{\end{fminipage}}
\def\bpic{\begin{picture}}              \def\epic{\end{picture}}
\def\bqun{\begin{quotation}}            \def\equn{\end{quotation}}
\def\bsl{\begin{slide}}                 \def\esl{\end{slide}}
\def\btabb{\begin{tabbing}}             \def\etabb{\end{tabbing}}
\def\btabl{\begin{table}}               \def\etabl{\end{table}}
\def\btablx{\begin{table*}}             \def\etablx{\end{table*}}
\def\btabu{\begin{tabular}}             \def\etabu{\end{tabular}}
\def\btabx{\begin{tabular*}}            \def\etabx{\end{tabular*}}
\def\bbib{\begin{thebibliography}{999}} \def\ebib{\end{thebibliography}}
\def\bver{\begin{verbatim}}             \def\ever{\end{verbatim}}
\def\bca{\begin{cases}}                  \def\eca{\end{cases}}
\def\brk{\begin{remark}}                  \def\erk{\end{remark}}

\def\cl#1{\centerline{#1}}

\def\MC{_{\scriptscriptstyle \text{LS}}}
\def\MCNM{_{\scriptscriptstyle \text{MNLS}}}
	
\begin{document}

\title{Fast 3D Synthetic Aperture Radar Imaging from Polarization-Diverse Measurements}
\author{Pierre~Minvielle, Pierre~Massaloux,
        and~Jean-Fran\c cois~Giovannelli
\thanks{P. Massaloux and P. Minvielle are with CEA, DAM, CESTA, F-33114 Le Barp, France.}
\thanks{J.-F. Giovannelli is with Univ. Bordeaux, IMS, UMR 5218, F-33400 Talence, France.}
}


\maketitle

\begin{abstract}
An innovative 3-D radar imaging technique is developed for fast and efficient identification and characterization of radar backscattering components of  complex objects, when the collected scattered field  is made of polarization-diverse measurements. In this context, all the polarimetric information seems irretrievably mixed. A  direct model, derived from a simple but original extension of the widespread ``multiple scattering model'' leads to a high dimensional linear inverse problem. It is solved by a fast dedicated imaging algorithm that performs to determine at a time three huge 3-D scatterer maps which correspond to HH, VV and HV  polarizations at emission and reception. It is applied successfully to various mock-ups and data sets collected from an accurate and dedicated 3D  spherical experimental layout that provides concentric polarization-diverse RCS measurements. 
\end{abstract}

\IEEEpeerreviewmaketitle

\section{Introduction}
\IEEEPARstart{S}{ynthetic} Aperture Radar (SAR) imaging is a widespread technique to make a map or image of the spatial distribution of reflectivity or backscattering of an object or scene from measurements of the scattered electric field. It is of great importance in many areas: land mapping, target recognition, environment monitoring, surveillance, nondestructive testing, etc. Radar imaging can also be used in an analysis purpose. For instance, to reduce the interference of wind turbine blades with air traffic control radars \cite{pinto2010stealth} or to characterize tree signatures for canopy monitoring \cite{fortuny1999three}. It is then  called "Radar Cross Section analysis" or ``RCS analysis''. Regarding the data, it is usually collected from a remote platform, for instance a satellite or an airborne. It can also be collected nearby, e.g. inside an indoor facility, if the object is small enough, for analysis imaging. Most of the time, radar imaging is one-dimensional (1-D) or two-dimensional (2-D), leading respectively to 1-D backscattering profiles or 2-D backscattering maps. Three-dimensional (3-D) radar images are known to be far more complicated and challenging to process \cite{mensa1991high}, beyond the required extensive signal processing. Indeed, if a two-dimensional (2D) backscattering image can be formed  by synthesizing an 1-D aperture with a wide-band radar, a three-dimensional (3D) backscattering image needs an entire 2D aperture. More precisely, in an indoor anechoic chamber, typical 2D aperture geometries are planar, spherical and cylindrical \cite{lopez20003}. Their physical implementations require many illumination viewpoints around the object. It requires to combine many object and/or antenna rotations, a time-consuming task while the measurement conditions may derive. Once the backscatter field data is recorded, it is processed in order to get 3D images of the target radar backscattering spatial distribution.  

Many techniques have been developed for 3D radar image formation, both in  SAR  where the radar platform is moving while the target stays motionless and  ISAR (Inverse SAR) where the target is moving while the radar platform radar is stationary. High resolution (HR) radar images are commonly  obtained by processing coherently the backscattered fields as a function of the frequency and the angle (i.e. object attitude relatively to the radar). Among 3D HR radar imaging techniques emphasized in \cite{lopez20003}, there are the Polar Format Algorithm (PFA), also known as Range-Doppler, the Range Migration Algorithm (RMA), its 2D version being also known as $k$-$\omega$ algorithm, and the Chirp Scaling Algorithm. Based on the polar nature of the frequency-domain backscatter data, 3D PFA is extensively used. Taking advantage of the processing that reduces in far-field condition to a Fourier synthesis problem, it is practically achieved via an interpolation, that reformats the data in the spatial frequency domain, and an inverse fast Fourier transform (FFT). Refer to \cite{mensa1991high} or \cite{ozdemir2012inverse}  for implementation details on 3D PFA, including the polar reformatting mapping technique. Regarding RMA, it comes from seismic engineering and geophysics. Based on 1-D Stolt interpolation and again FFT through an approximation with the method of stationary phase (MSP), RMA is able to compensate completely a potential wavefront curvature. The last one, i.e. the Chirp Scaling Algorithm, is widely used in airborne SAR. Besides, let us mention space-time domain methods, such as the 3D Backprojection (BP) algorithm.  Also known as Time-Domain Correlation, this tomographic reconstruction is achieved by coherent summation and stems from  the projection-slice theorem; see for example \cite{knaell1995radar}. 

Back to the far-field condition of PFA, it can be relaxed even if it must be stressed that 3D near-field radar imaging remains computationally expensive \cite{mensa1991high,fortuny1999three}. Let us mention for example \cite{fortuny98} that develops a near-field focusing function that accounts for wavefront curvature and propagation loss; it is employed in \cite{fortuny1999three} for RCS  analysis of large trees inside an anechoic chamber. Another condition, generally implicitly required in radar imaging, is the so-called "small bandwidth  small angle" \cite{ozdemir2012inverse}. This assumption means that the data is collected for a bounded excursion of the observation angles and frequency where the nature of the wave-object interaction is unchanged. It is closely related to the validity domain of the standard "multiple scatterer model" where the object is represented by a collection of coherently illuminated point scatterers \cite{cheney2009fundamentals}, the properties of which do not vary in the limited frequency and angle excursion band.

In this article, we present a fast and efficient 3D non conventional radar imaging technique for identification and characterization of radar reflectivity/backscattering components of complex objects, in the context of  RCS or signature analysis. It goes far beyond the previously proposed ad hoc localization processing technique, the limitations of which are described in \cite{Minvielle}. Unlike \cite{fortuny1999three}, our spherical measurement set-up, designed and developed at CEA, is especially appropriate for measuring  small targets of lower RCS.  The specific feature is that the collected scattered field data is made of polarization-diverse measurements: the electric field varies both at emission and reception during the acquisition, forming concentric circles with a singularity in the main direction.  In a way, the issue is related to polarimetric imaging and to what is often called imaging radar polarimeter \cite{zebker1991imaging}. SAR polarimetry is a widely used technique for measuring and identifying polarimetric properties of scatterers \cite{moreira2013tutorial,Lee}. Many works have been achieved in HR polarimetric characterization or decomposition, especially for automatic target recognition \cite{wang2001characterization,martorella2009automatic}, in order to determine the full scatterer matrix of a scatterer and get a relevant information on the shape, orientation or dielectric properties.  Here, our singular set-up implies that all the polarimetric information is mixed. To overcome the issue, an original approach is proposed and developed. It is based on a suitable extension of the multiple scattering model that leads to  a high dimensional linear direct problem. This inverse problem is solved by a fast regularization algorithm that manages to determine at a time three huge 3D scatterer maps.

The article is organized as follows. First, in section \ref{Section_Problem}, a  general description of the  polarization-diverse measurement problem is made. In section \ref{Section_Imaging} is dedicated to the 3D radar imaging approach. After a brief introduction to classical radar imaging and the multiple scatterer model, an extension is developed. Inspired by HR polarimetric characterization, it leads to the expression of the direct model and to the proposition of a fast regularized inversion method. In  section \ref{Section_Application}, the 3D spherical RCS set-up is presented: it is able to accurately acquire polarization-diverse measurements. After a few details on the associated specific 3D radar image processing, results are presented and discussed for various mock-ups.  Finally, conclusions are summarized in section~\ref{Section_Conclusion}.

\section{Problem formulation} \label{Section_Problem}

\subsection*{Standard RCS Synthetic Aperture Radar}
A monostatic radar illuminates an object  with a quasi-planar monochromatic continuous wave (CW) of given frequency $f$. The incident electric field is vector $\mathbf{E}^I$ of complex amplitude $\mathrm{E}^I$. The object backscatters a CW with the same frequency. The scattered electric field is vector $\mathbf{E}^S$, of complex amplitude $\mathrm{E}^S$. The complex scattering coefficient $\mathcal{S}$ quantifies the whole object-EM wave interaction; it indicates the wave change in amplitude and phase. Note that it is directly linked to the Radar Cross Section. They can be defined in far field condition by (see \cite{Knott} for more details):
 \begin{equation} \label{def_RCS}
\mathcal{S}= \lim_{R \rightarrow \infty} 2\sqrt{\pi} R \frac{\mathrm{E}^S}{\mathrm{E}^I} \mbox{,} \quad \mathrm{RCS}=|\mathcal{S}|^2
\end{equation}
where $R$ is the radar distance at which scattering is observed.  The complex scattering coefficient $\mathcal{S}$ can be measured with an appropriate  instrumentation system (antenna, network analyzers, etc.) and a calibration process. The measurements can be repeated for different wave frequencies, which is usually called "Stepped Frequency Continuous Wave"  (SFCW) acquisition mode. They can be repeated also for different incidence angles, by rotating the object or/and the antenna. Finally, it leads to a sequence of $M$ measured complex scattering coefficients $\{\mathcal{S}_1,\mathcal{S}_2,\cdots,\mathcal{S}_M\}$. Considering all these successive measurements, it is generally assumed in HR radar imaging that the wave-object interaction nature is unaltered. In particular, the polarization configuration remains stable, i.e. the electric field orientation relatively to the target is unchanged.

\begin{figure}[!ht]
\centering \includegraphics[width=0.3\textwidth]{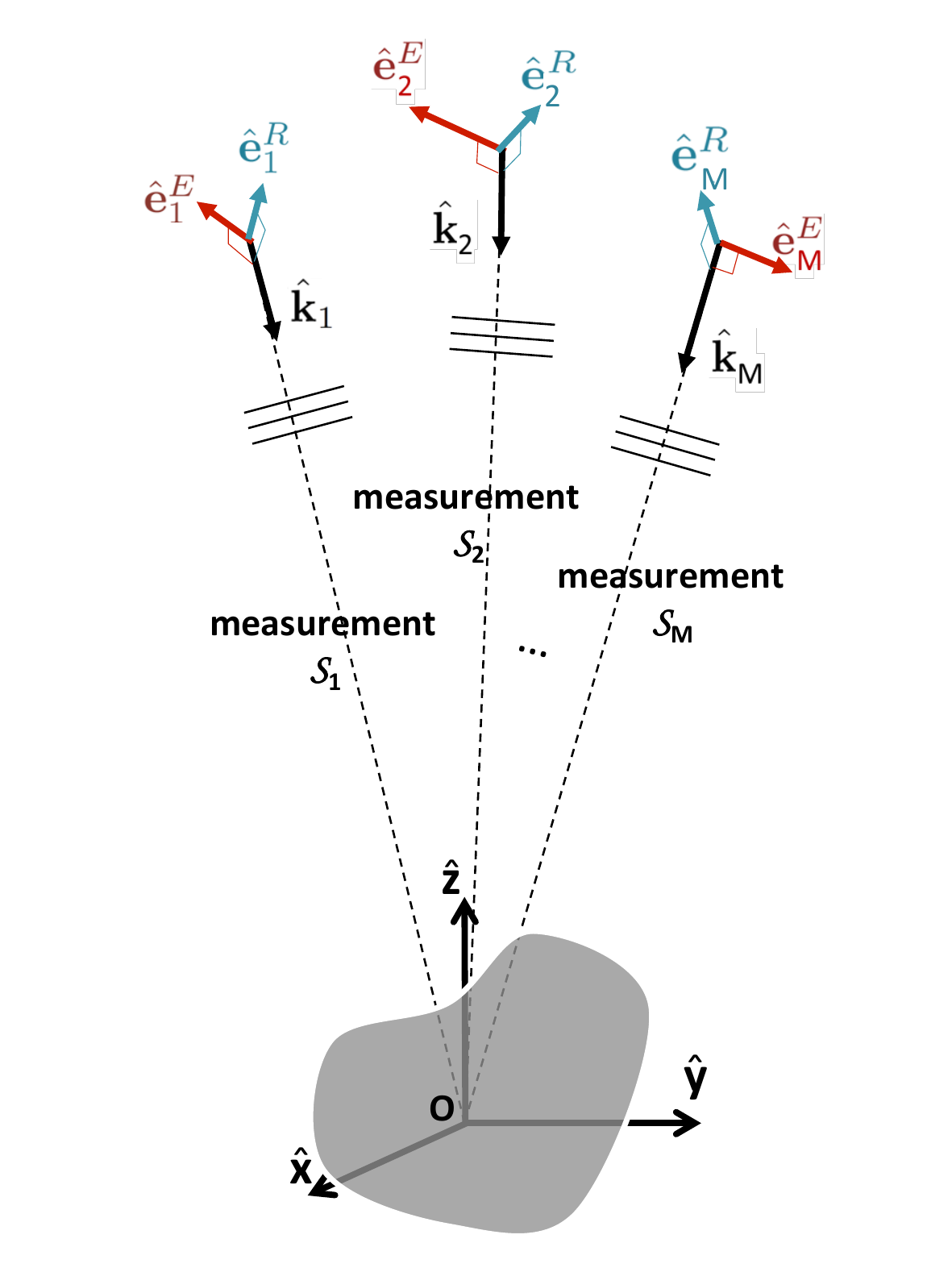} \caption{Polarization-diverse radar imaging acquisition at successive wave frequencies $\{f_1,f_2,\cdots,f_M\}$} \label{problem}
\end{figure}

\subsection*{Polarization-diverse measurements}
Far from above standard RCS SAR, we consider afterwards the singular but relevant problem of a monostatic acquisition composed of $M$ polarization-diverse measurements. It is represented in figure \ref{problem}. Our goal is to analysis the scattering from $O\hat{\mathbf{z}}$ viewpoint. In this singular set-up, each measurement is associated with a different linear polarization, i.e. a different electric field direction.  Let us focus on the $i$-th measurement at wave frequency $f_i$. It is described by the wave vector $\mathbf{k}_i$, i.e. its direction (unit vector $\hat{\mathbf{k}}_i$) and its modulus or frequency wavenumber $\mathrm{k}_i=2\pi f_i/c$ (where $c$ denotes the light speed), by the polarization direction at emission (unit Jones vector $\hat{\mathbf{e}}^E_i$) and by the polarization direction at reception (unit Jones vector $\hat{\mathbf{e}}^R_i$). The incident electric field $\mathbf{E}^I_i$ is collinear to  $\hat{\mathbf{e}}^E_i$ while $\hat{\mathbf{k}}_i \wedge \hat{\mathbf{e}}^E_i$ indicates the direction of the incident magnetic field.  Finally, the polarization-diverse measurement is defined, i.e. both microwave emission and reception, by the successive frequencies $\{f_1,f_2,\cdots,f_M\}$ and the successive  corresponding directional triplets $\{(\hat{\mathbf{k}}_1,\hat{\mathbf{e}}^E_1,\hat{\mathbf{e}}^R_1),(\hat{\mathbf{k}}_2,\hat{\mathbf{e}}^E_2,\hat{\mathbf{e}}^R_2),\cdots,(\hat{\mathbf{k}}_M,\hat{\mathbf{e}}^E_M,\hat{\mathbf{e}}^R_M)\}$.

Consequently, the polarization-diverse measurement 3D radar imaging problem consists in determining 3D scatterer maps from $O\hat{\mathbf{z}}$ radar viewpoint, based on this sequence of polarization-dependent measurements. Compared to classical radar imaging, the main issue is linked with the variation of polarization, i.e. the rotation of both the electric and magnetic fields. Indeed, the classical multiple isotropic scatterer model on which radar imaging usually relies is no longer valid. Whereas the backscatter data is usually recorded separately for each polarization at emission and reception (i.e. HH, VV and HV), here all the information is mixed up. Besides, the huge number of unknown quantities needs to be processed efficiently.
\section{Fast 3D radar imaging with polarization-diverse measurements} \label{Section_Imaging}

\subsection{Overview of classical 3D HR radar imaging and the multiple scatterer model}
Let us briefly introduce classical HR radar imaging \cite{mensa1991high} and first the so-called ubiquitous "multiple scatterer" model on which it is based. The multiple scatterer (MS) model lies on the interpretation of  the target-electromagnetic wave interaction by a collection of coherent illuminated and localized point scatterers \cite{cheney2009fundamentals}. Related to the high frequency behavior of EM scattering from geometrically complex bodies, the MS model can be derived from Maxwell's equations by high frequency approximations \cite{cheney2009fundamentals}. It must be stressed that it requires several restrictive conditions and does not explicitly account for various EM phenomena: multiple scattering, creeping waves, etc. Even so, the MS model is practically and intensively used in radar interpretation and analysis, far beyond the validity of its assumptions. It works pretty well in the far field  as long as there is a critical look at its output where possible artifacts may occur \cite{cheney2009fundamentals}, mainly related to the above mentioned EM phenomena. 

\subsubsection*{MS model} In a monostatic radar context with far field/planar monochromatic wave (with wave vector $\mathbf{k}_i$),  the MS model considers a target made of $N$ elementary isotropic scatterers. They are located at points $\mathbf{r}_n$ ($n=1,\cdots,N$), of coordinates $(x_n,y_n,z_n)$ in the target coordinate system $(O\hat{\mathbf{x}}\hat{\mathbf{y}}\hat{\mathbf{z}})$. Each scatterer is defined by its complex scattering coefficient or power $s_n$ ($n=1,\cdots,N$), that quantifies the relationship between the incident and scattered field amplitudes ($\mathrm{E}^I$ and $\mathrm{E}^S_n$)\footnote{It is assumed that the incident scattered field amplitude $\mathrm{E}^I$ does not depend on $n$.}, resulting from the elementary wave-scatterer interaction. Considering a large enough radar-target distance $R$, it is possible to derive from (\ref{def_RCS}):
\begin{equation} \label{MS_1}
\mathrm{E}^S_n \approx\frac{1}{2\sqrt{\pi} |R \cdot \hat{\mathbf{u}}_R +\mathbf{r}_n|} (s_n\cdot e^{-2j\mathbf{k}_i\cdot \mathbf{r}_n}) \cdot \mathrm{E}^I
\end{equation} 
where $\hat{\mathbf{u}}_R$ is the unit vector corresponding to radar-$O$ direction  in $(O\hat{\mathbf{x}}\hat{\mathbf{y}}\hat{\mathbf{z}})$, so that $|R \cdot \hat{\mathbf{u}}_R +\mathbf{r}_n|$ is the distance between the radar and the current scatterer $n$. In the phase shift, the term $2 \mathbf{k}_i\cdot\mathbf{r}_n$ accounts for the back and forth delays. 
 
In the MS model, $\mathrm{E}^S=\sum_{n=1}^{N} \mathrm{E}^S_n$, i.e. the received echo or scattered signal is ideally formulated as a coherent superposition of elementary echoes \cite{cheney2009fundamentals}, neglecting multiple scattering terms. That leads to the following scattering coefficient:
\begin{equation} \label{ModPb0}
 \mathcal{S}_i = 2\sqrt{\pi} R \frac{\mathrm{E}^S}{\mathrm{E}^I} = \sum_{n=1}^{N} \frac{R}{|R \cdot \hat{\mathbf{u}}_R +\mathbf{r}_n|} s_n \cdot e^{-2j\mathbf{k}_i\cdot\mathbf{r}_n}
\end{equation} 

Since the ratio $\frac{R}{|R \cdot \hat{\mathbf{u}}_R +\mathbf{r}_n|}= \frac{1}{|\hat{\mathbf{u}}_R +\mathbf{r}_n / R|}\approx 1$, the MS model reduces to:
\begin{equation} \label{EquModPb}
 \mathcal{S}_i=\sum_{n=1}^{N} s_n \cdot e^{-2j\mathbf{k}_i\cdot\mathbf{r}_n}
\end{equation} 

The $N$ complex scattering coefficients $s_n$  associated to the multiple scatterers determine the system response. They do not depend on the wave vector $\mathbf{k}_i$, i.e. its amplitude and direction. 

\subsubsection*{Typical application} 
\begin{figure}[ht!] \centering 
\includegraphics*[width=0.5\textwidth, clip=true,viewport=5cm 2cm 23cm 17.5cm] {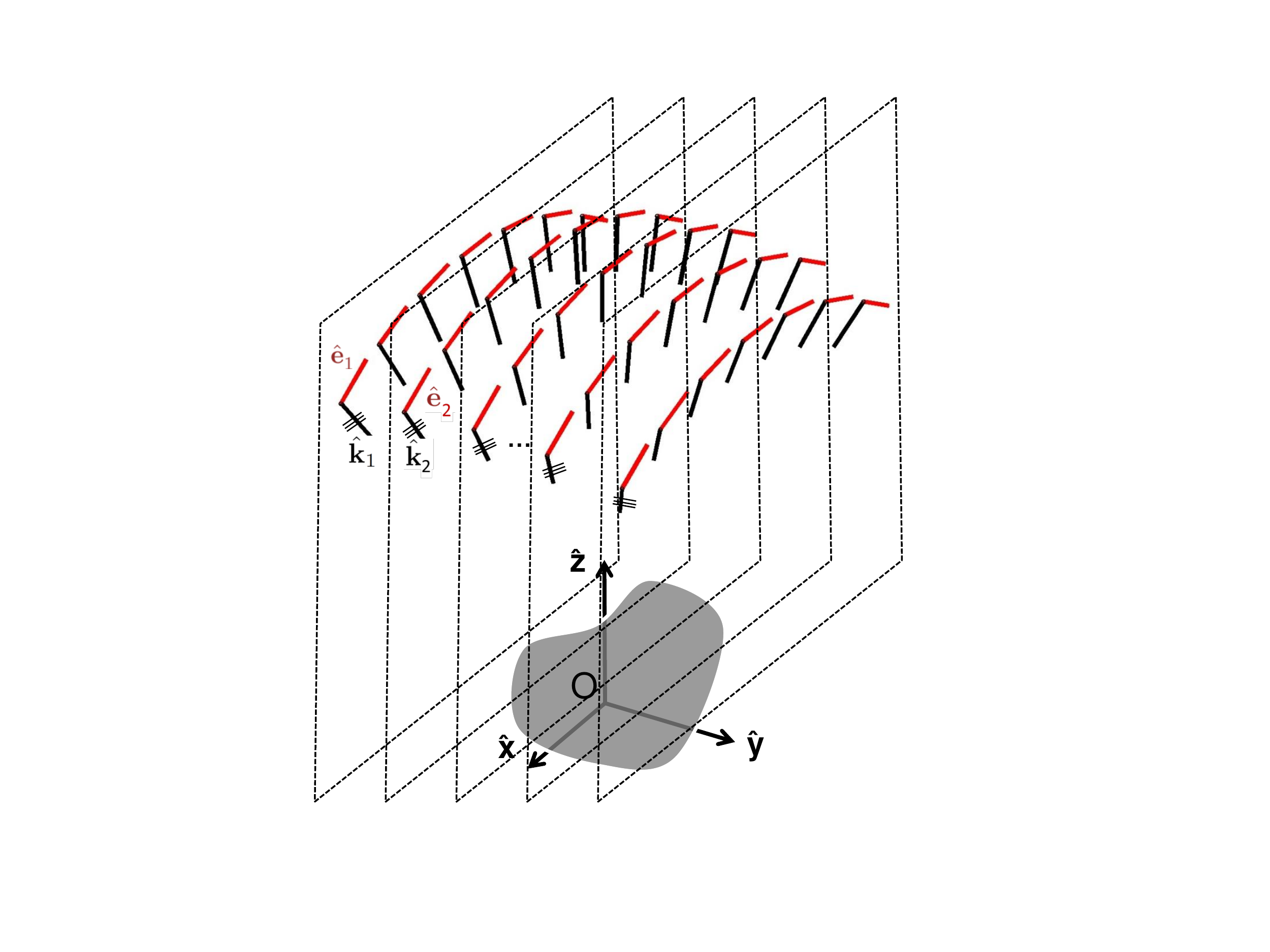}
\caption{Classical spherical radar imaging with $\hat{\mathbf{e}}^E_i=\hat{\mathbf{e}}^R_i=\hat{\mathbf{e}}_i$} \label{Imag3D} 
\end{figure}

Consider the typical acquisition of Fig.~\ref{Imag3D} that can be encountered in spherical 3D radar imaging. It requires a sequence of $M$ viewpoints around $O\hat{\mathbf{z}}$.  Considering that the \emph{i-th} wave direction is defined by the standard spherical angles $(\vartheta_i,\Phi_i)$, the successive wave vector directions $\hat{\mathbf{k}}_i$ are given by:
\begin{equation*} 
\hat{\mathbf{k}}_i=[-\cos\vartheta_i\sin\Phi_i  \quad -\sin\vartheta_i \quad -\cos\vartheta_i\cos\Phi_i ]\T
\end{equation*} 

It must be noticed that in such an acquisition the electric field varies slightly as long as the variations of $(\vartheta_i,\Phi_i)$ are limited. This is also called the "small angle" condition. Moreover,  the electric field stays in the same plane, as shown in Fig.~\ref{Imag3D}, and the polarization remains more or less constant. Another condition is required: it is called the "small bandwidth" condition. It implies that the target-wave interaction does not significantly vary within the limited frequency variation domain. If these conditions are fulfilled, each complex scattering coefficient $s_n$ is considered constant for all the measurements. The MS model (\ref{EquModPb}) can be formulated for each measurement $i$ (for $i=1,\cdots,M$) as: 
\begin{equation*} 
 \mathcal{S}_i=\sum_{n=1}^{N} s_n \cdot e^{2j \mathrm{k}_i\cdot (\sin\vartheta_i\cos\Phi_i x_n+\sin\vartheta_i\sin\Phi_i y_n+\cos\vartheta_i z_n)}
\end{equation*}

Remark that the MS model can be formulated as:
\begin{equation} \label{ModPBCont}
 \mathcal{S}_i=\int \mathcal{I}(\mathbf{r}) \cdot e^{-2j\mathbf{k}_i\cdot\mathbf{r}} \cdot d\mathbf{r} + \varepsilon
\end{equation} 
where  $\mathcal{I}(\mathbf{r})=\sum_{n=1}^{N}s_n \delta(\mathbf{r}-\mathbf{r}_n)$ is the spatial density of scatterers and $\delta(\cdot)$ is the Dirac function. The additive term $\varepsilon$ accounts for noise and uncertainty, due to interfering echoes generated by the environment and MS model limitations. It is generally considered as white and Gaussian (centered, complex and circular). 

\subsubsection*{Radar imaging} Radar imaging processing is an inverse problem. For a given radar viewpoint (e.g. $O\hat{\mathbf{z}}$), it consists of the determination the complex 3D image or map $\mathcal{I}$, i.e. the complex amplitudes $s_i$ at points $\mathbf{r}_i$. The above mentioned "small  bandwidth small angle" condition is assumed. It must be stressed that it requires that the wave polarization remains constant so that the target-wave interaction does not vary from one measurement to another. 

The Polar Format Algorithm (PFA), formerly introduced, is based on  (\ref{ModPBCont}) which is the discrete Fourier transform (DFT) of the sought image $\mathcal{I}$. The computation of $\mathcal{I}$ is achieved via the inverse Fourier transform of the measured hologram, $\mathcal{S}_j$ (for $j=1,\cdots,M$). In actual practice, the limited and discrete acquisition in  angles and in frequencies must be taken into account. For a regular acquisition in angle and frequency, an interpolating pre-processing step of regridding is performed before an inverse 3D Fast Fourier transform. 

\subsection{Extension to the multiple scatterer model}
Considering the polarization-diverse acquisition of Fig.~\ref{problem},  the MS model is no longer valid and as a consequence conventional radar imaging processing can not be applied. Actually, the complex scattering coefficients $s_n$ ($n=1,\cdots,N$) can no longer be considered constant.  The electromagnetic interaction strongly depends on the electric field direction;  consider for example the scattering with the edges of an object. To overcome this issue, we next develop an adaptation of the MS model in this varying polarization context that will latter support the imaging inversion process. 

\subsubsection*{General formulation}
Remind that the purpose is to achieve radar imaging analysis from $O\hat{\mathbf{z}}$ radar viewpoint. Let us consider the \emph{i-th} acquisition of Fig.~\ref{problem}, around $O\hat{\mathbf{z}}$. Then, the basic idea is to consider that each elementary isotropic scatterer  is associated with not only a complex scattering coefficient $s_n$ but with a full polarization scattering matrix $\mathbf{S}_n$:
\begin{equation} 
\mathbf{S}_n=
\left[
\begin{array}{lr}   s_n^{\textrm{xx}}  & s_n^{\textrm{xy}}  \\ s_n^{\textrm{xy}} & s_n^{\textrm{yy}}  \end{array} 
\right]
\end{equation} 
More exactly known as the \emph{radar cross section matrix} \cite{Knott_m},  $\mathbf{S}_n$ is defined in the target reference frame $O\hat{\mathbf{x}}\hat{\mathbf{y}}$, associated with an illumination direction collinear to vector $\hat{\mathbf{z}}$. It quantifies the amplitude and polarization of the scattered wave for an arbitrary polarization of the incident wave. Note that $s_n^{\textrm{xy}}=s_n^{\textrm{yx}}$ due to the reciprocity theorem for a monostatic case \cite{Lee}, so that the scattering matrix $\mathbf{S}_n$ is completely defined by the three complex coefficients: $s_n^{\textrm{xx}}$, $s_n^{\textrm{yy}}$ and $s_n^{\textrm{xy}}$. Next, we derive an extension to the MS model. For  each illumination,  it introduces the appropriate combination of scattering matrix terms. 

Introducing polarization, the former general definition (\ref{def_RCS}) of the complex scattering coefficient can be reformulated by:
 \begin{equation} \label{def_RCS_2}
\mathcal{S} =  \lim_{R \rightarrow \infty} 2\sqrt{\pi} R \frac{\mathrm{T}^S}{\mathrm{E}^I}
\end{equation}
where $\hat{\mathbf{e}}_R$ is a unit vector aligned along the electric polarization at reception and the scalar product $\mathrm{T}^S=\hat{\mathbf{e}}_R \cdot \mathbf{E}^S$ corresponds to the transmitted electric field. Refer to \cite{Knott} for more details. 

\begin{figure}[ht!] \centering 
\includegraphics*[width=0.3\textwidth, clip=true,viewport=0cm 0cm 14cm 15.5cm] {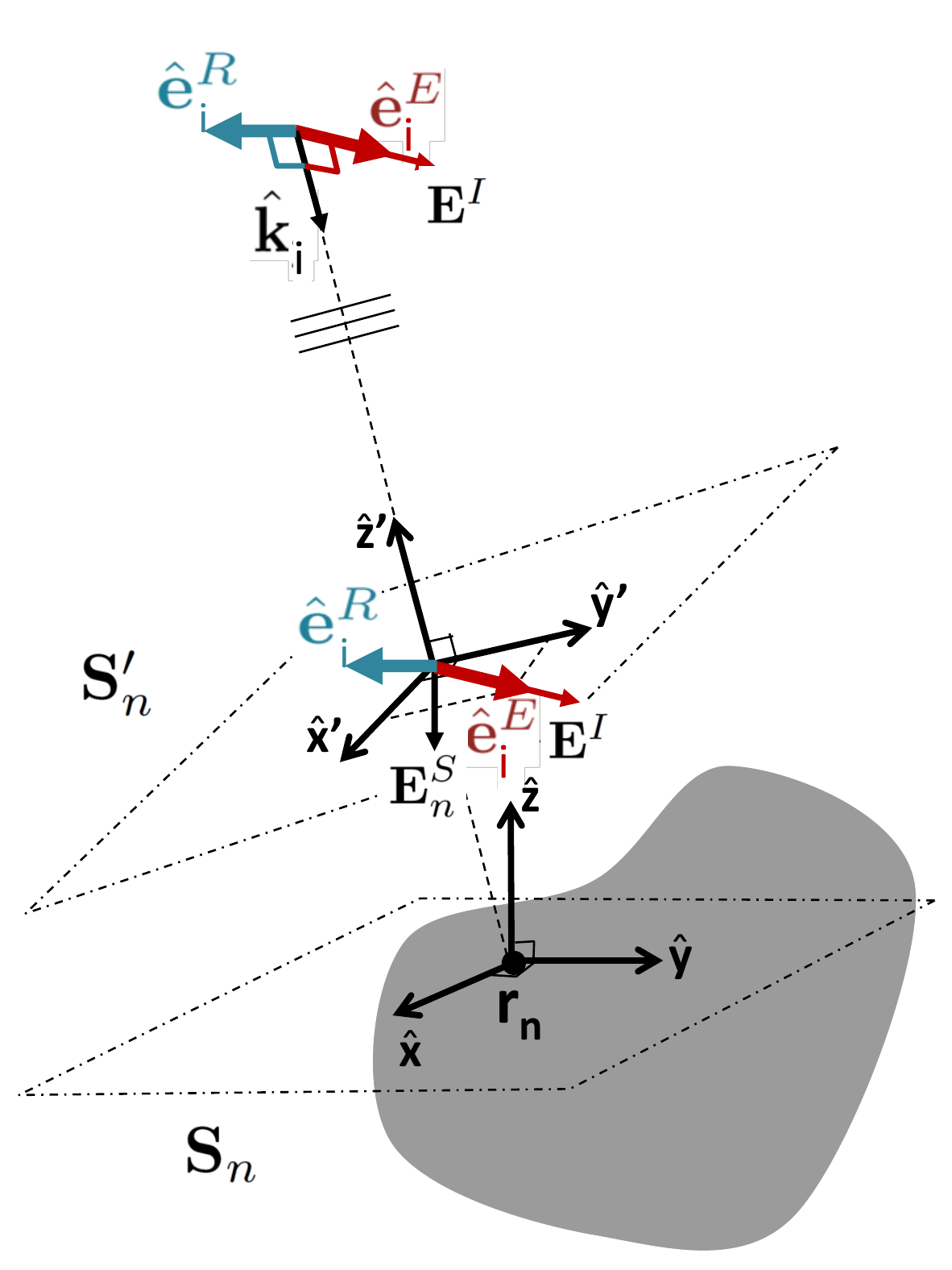}
\caption{Interaction of an EM wave with a point scatterer} \label{extension} 
\end{figure}

The polarization configuration of the \emph{n-th} scatterer is represented in Fig.~\ref{extension}. For the $i$-th acquisition, the scattering field vector $\mathbf{E}^S_n$ resulting from the interaction between the incident wave and the \emph{n-th} elementary isotropic scatterer is given in  $(O\hat{\mathbf{x}}^\prime\hat{\mathbf{y}}^\prime)$ by (for a large target-radar distance $R$):
\begin{eqnarray}
{\mathbf{E}^S_n}_{ (\hat{\mathbf{x}}^\prime,\hat{\mathbf{y}}^\prime)} &\approx&\frac{e^{-2j\mathbf{k}_i\cdot\mathbf{r}_n}}{2\sqrt{\pi} |R \cdot \hat{\mathbf{u}}_R +\mathbf{r}_n|} ({\mathbf{S}^\prime_n(i) \cdot \mathbf{E}^I_i}_{ (\hat{\mathbf{x}}^\prime,\hat{\mathbf{y}}^\prime)})  \nonumber\\ &=& \frac{e^{-2j\mathbf{k}_i\cdot\mathbf{r}_n}}{2\sqrt{\pi} |R \cdot \hat{\mathbf{u}}_R +\mathbf{r}_n|} \left(\mathbf{S}^\prime_n(i) \cdot
\left[
\begin{array}{c} 
\mathbf{E}^I_i \cdot \hat{\mathbf{x}}^\prime \\ 
\mathbf{E}^I_i \cdot \hat{\mathbf{y}}^\prime 
\end{array}
\right]\right)  \nonumber
\end{eqnarray} 
where $\mathbf{S}^\prime_n(i)$ is the scattering matrix in the reference frame $O\hat{\mathbf{x}}^\prime\hat{\mathbf{y}}^\prime$, normal to the current direction of propagation  $\hat{\mathbf{k}}_i$. Note that there is no component of $\mathbf{E}^S_n$ and $\mathbf{E}^I_i$ in the direction of $\hat{\mathbf{k}}_i$.

Let denote $\hat{\mathbf{e}}^E_i$ the unit Jones vector at emission and $\hat{\mathbf{e}}^R_i$ the unit Jones vector associated with the receiving linear polarized antenna. The radar received voltage is proportional to the transmitted electric field: $\mathrm{T}^S_n= [\hat{\mathbf{e}}^R_{i_{(\hat{\mathbf{x}}^\prime,\hat{\mathbf{y}}^\prime)}}]\T \cdot {\mathbf{E}^S_n}_{ (\hat{\mathbf{x}}^\prime,\hat{\mathbf{y}}^\prime)}$, so that:
\begin{equation*}
\mathrm{T}^S_n  = \frac{e^{-2j\mathbf{k}_i\cdot\mathbf{r}_n}}{2\sqrt{\pi} |R \cdot \hat{\mathbf{u}}_R +\mathbf{r}_n|}
\left[
\begin{array}{c} 
\hat{\mathbf{e}}^R_i \cdot \hat{\mathbf{x}}^\prime \\ 
\hat{\mathbf{e}}^R_i\cdot \hat{\mathbf{y}}^\prime
\end{array}
\right]\T 
\cdot \mathbf{S}^\prime_n(i) \cdot
\left[
\begin{array}{c} 
\mathbf{E}^I_i \cdot \hat{\mathbf{x}}^\prime \\ 
\mathbf{E}^I_i \cdot \hat{\mathbf{y}}^\prime 
\end{array}
\right]  
\end{equation*} 
It comes to:
\begin{equation*}
\mathrm{T}^S_n=\frac{\mathrm{E}^I e^{-2j\mathbf{k} \cdot \mathbf{r}_n}}{2\sqrt{\pi} |R \cdot \hat{\mathbf{u}}_R +\mathbf{r}_n|}
\left[
\begin{array}{c} 
\hat{\mathbf{e}}^R_i \cdot \hat{\mathbf{x}}^\prime \\ 
\hat{\mathbf{e}}^R_i\cdot \hat{\mathbf{y}}^\prime
\end{array}
\right]\T \cdot \mathbf{S}^\prime_n(i) 
 \cdot
\left[
\begin{array}{c} 
\hat{\mathbf{e}}^E_i\cdot \hat{\mathbf{x}}^\prime \\ 
\hat{\mathbf{e}}^E_i \cdot \hat{\mathbf{y}}^\prime 
\end{array}
\right]  
\end{equation*} 

Similarly to the standard MS model, the received echo $\mathrm{T}^S$ can be simply formulated as a coherent superposition of the elementary echoes: $\mathrm{T}^S=\sum_{n=1}^{N} \mathrm{T}^S_n$. Thus, the complex scattering coefficient scattering coefficient is:

\begin{equation} \label{MS_E_0}
\mathcal{S}_i=2\sqrt{\pi} R \frac{\mathrm{T}^S}{\mathrm{E}^I}=2\sqrt{\pi} R \frac{\sum_{n=1}^{N} \mathrm{T}^S_n}{\mathrm{E}^I}
\end{equation} 

In the far field  ($\mathbf{r}_n/R \approx 0$), it is then straightforward to derive the following extended MS model for polarization-diverse measurement, incorporating noise:
\begin{equation} \label{Eq:MS_ext}
 \mathcal{S}_i=\sum_{n=1}^{N} s^\star_n(i) \cdot e^{-2j\mathbf{k}_i\cdot\mathbf{r}_n} + \varepsilon_i
\end{equation} 
\begin{equation} \label{s_n_star}
\mbox{with} \quad s^\star_n(i)=[\hat{\mathbf{e}}^R_i\cdot \hat{\mathbf{x}}^\prime \quad \hat{\mathbf{e}}^R_i \cdot \hat{\mathbf{y}}^\prime] \cdot \mathbf{S}^\prime_n(i) \cdot
\left[
\begin{array}{c} 
\hat{\mathbf{e}}^E_i\cdot \hat{\mathbf{x}}^\prime \\ 
\hat{\mathbf{e}}^E_i \cdot \hat{\mathbf{y}}^\prime 
\end{array}
\right] 
\end{equation} 

As previously with the MS model about the scattering coefficients $s_n$, a "small bandwidth small angle" will be further assumed about the scattering matrices. Due to a limited acquisition domain, in angle (around $O\hat{\mathbf{z}}$) and frequency, the scattering matrices are supposed to be nearly constant: $\mathbf{S}^\prime_n(i) \approx \mathbf{S}_n$ for $i=1,\cdots,M$. Former expression (\ref{s_n_star}) leads to:  
\begin{equation} \label{Eq:MS_ext_rho}
s^\star_n(i)\approx[\hat{\mathbf{e}}^R_i\cdot \hat{\mathbf{x}}^\prime \quad \hat{\mathbf{e}}^R_i \cdot \hat{\mathbf{y}}^\prime] \cdot \mathbf{S}_n \cdot
\left[
\begin{array}{c} 
\hat{\mathbf{e}}^E_i\cdot \hat{\mathbf{x}}^\prime \\ 
\hat{\mathbf{e}}^E_i \cdot \hat{\mathbf{y}}^\prime 
\end{array}
\right] 
\end{equation} 
Basically, this expression provides the complex scattering coefficients or powers $s^\star_n$ of the MS model that are adapted to the \emph{i-th} configuration. Note again that it depends on the relative target attitude towards the emitting and receiving antenna. In classical radar imaging, the directions of $\hat{\mathbf{e}}^E_i$ and $\hat{\mathbf{e}}^R_i$ remain stable, selecting constantly a term of the scattering matrix $\mathbf{S}_n$  or a fixed linear combination of them. In this varying polarization context, they are all "mixed": (\ref{Eq:MS_ext}) and (\ref{Eq:MS_ext_rho}) show that each acquisition is to select a different combination of the scattering matrix $\mathbf{S}_n$. Next, the extended MS model is detailed for a specific set-up.

\subsubsection*{Specific form for concentric polarization-diverse measurements} \label{sec_polardiv}
\begin{figure}[ht!] \centering 
\includegraphics*[width=0.5\textwidth, clip=true,viewport=5cm 3cm 23cm 14cm] {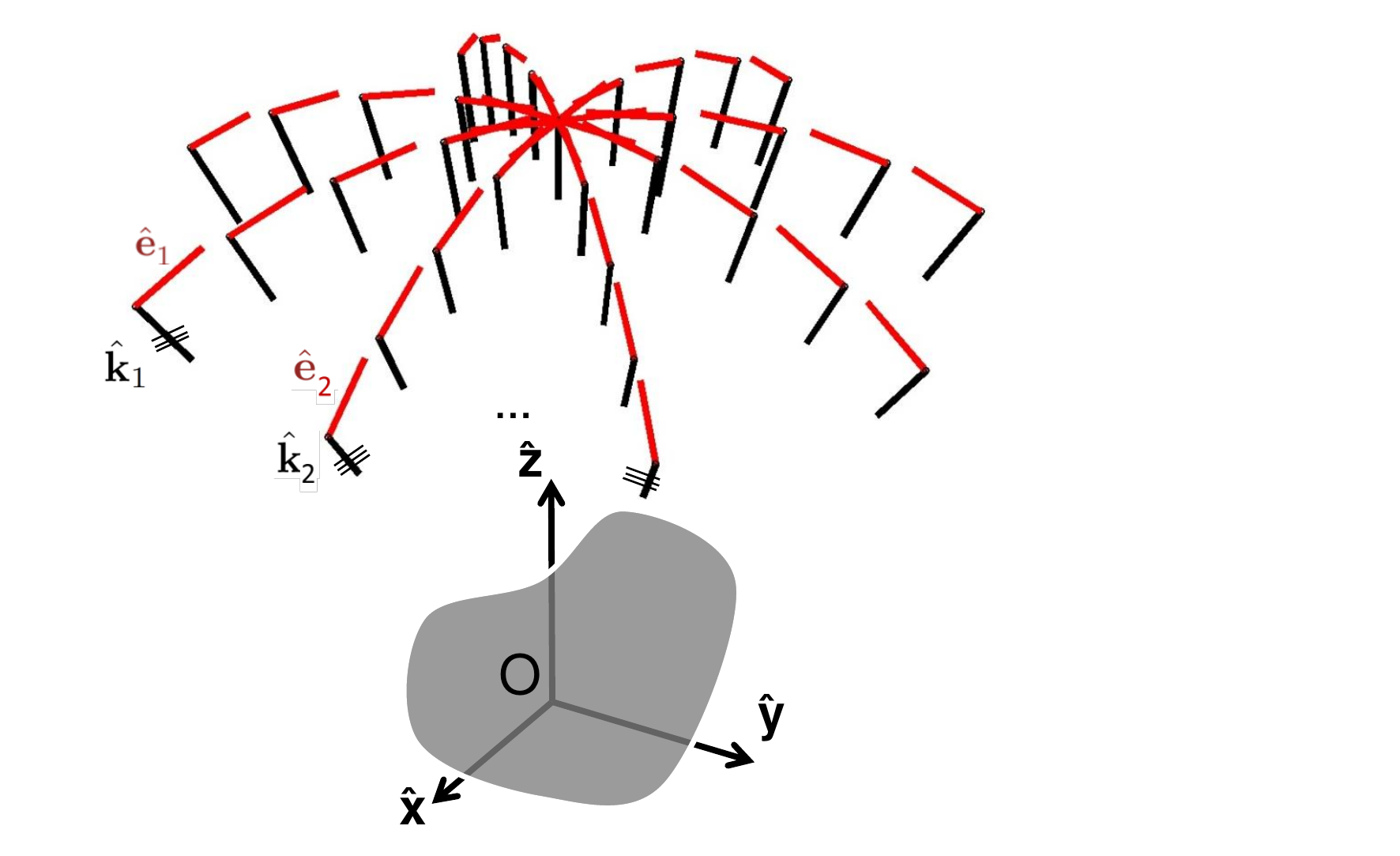}
\caption{Concentric Polarization-Diverse Radar Imaging Acquisition} \label{Imag3Dsing} 
\end{figure}

Let us consider the concentric acquisition, with the successive wave and electric field vectors represented in Fig.~\ref{Imag3Dsing}. It stems from a spherical 3D RCS set-up that will be further detailed in section \ref{Section_Application}. Firstly, note the following sort of singularity: the vectors spin around axis $\hat{\mathbf{z}}$ and there are various acquisitions from wave vectors colinear to $\hat{\mathbf{z}}$ while electric fields differ. Note that it induces different scattered fields, it goes against the classical MS model with scalars $s_n$.   

The \emph{i-th} measurement is given by $(\theta_i,\varphi_i,f_i)$, for $i=1,\cdots,M$. The azimuth $\theta_i$ and the roll $\varphi_i$ correspond to two rotation angles (the roll rotation is defined around $\hat{\mathbf{z}}$, see section \ref{Section_Application} for details), that defines the directions of the wave vector direction $\hat{\mathbf{k}_i}$ as well as  the emitting/receiving linear polarization:

\begin{equation*}
 \hat{\mathbf{k}}_i=\left[
			\begin{array}{c}
			 -\sin\theta_i \cos\varphi_i \\
			 -\sin\theta_i\sin\varphi_i  \\
			 -\cos\theta_i
			\end{array}
		\right]_{(\hat{\mathbf{x}},\hat{\mathbf{y}},\hat{\mathbf{z}})}
\end{equation*} 
\begin{equation*}
\hat{\mathbf{e}}^E_i=\left[
			\begin{array}{c}
			 -\cos\theta_i \cos\varphi_i \\
			 -\cos\theta_i \sin\varphi_i \\
			 \sin\theta_i
			\end{array}
		\right]_{(\hat{\mathbf{x}},\hat{\mathbf{y}},\hat{\mathbf{z}})}
\mbox{ or }
\left[
			\begin{array}{c}
			 -\sin\varphi_i \\
			 \cos\varphi_i\\
			 0
			\end{array}
		\right]_{(\hat{\mathbf{x}},\hat{\mathbf{y}},\hat{\mathbf{z}})}
\end{equation*} 
depending on the emitting and receiving linear polarized mode of the antenna (respectively H or V), and so for $\hat{\mathbf{e}}^R_i$.

Afterwards, we determine the specific forms of $(\hat{\mathbf{e}}^R_i\cdot \hat{\mathbf{x}}^\prime, \hat{\mathbf{e}}^R_i \cdot \hat{\mathbf{y}}^\prime)$ and $(\hat{\mathbf{e}}^E_i\cdot \hat{\mathbf{x}}^\prime, \hat{\mathbf{e}}^E_i \cdot \hat{\mathbf{y}}^\prime)$ in order to express $s^\star_n(i)$ for concentric acquisition. First of all, let us define the intermediate coordinate system  $(\hat{\mathbf{x}}^\prime_0,\hat{\mathbf{y}}^\prime_0,\hat{\mathbf{z}}^\prime_0)$:
\begin{eqnarray}
\hat{\mathbf{z}}^\prime_0&=&-\hat{\mathbf{k}_i} \\
 \hat{\mathbf{y}}^\prime_0&=& \frac{\hat{\mathbf{z}}^\prime_0 \wedge \hat{\mathbf{z}}}{\parallel \hat{\mathbf{z}}^\prime_0 \wedge \hat{\mathbf{z}}\parallel} =\left[
			\begin{array}{c}
			 -\sin\varphi_i \\
			 \cos\varphi_i \\
			 0
			\end{array}
		\right]_{(\hat{\mathbf{x}},\hat{\mathbf{y}},\hat{\mathbf{z}})} \\
 \hat{\mathbf{x}}^\prime_0&=& \hat{\mathbf{y}}^\prime_0 \wedge \hat{\mathbf{z}}^\prime_0 =\left[
			\begin{array}{c}
			 \cos\theta_i \cos\varphi_i \\
			 \cos\theta_i \sin\varphi_i  \\
			 -\sin\theta_i
			\end{array}
		\right]_{(\hat{\mathbf{x}},\hat{\mathbf{y}},\hat{\mathbf{z}})}
\end{eqnarray}

The unit Jones vector at emission $\hat{\mathbf{e}}^E_i$, and so at reception $\hat{\mathbf{e}}^R_i$, can be expressed in $(\hat{\mathbf{x}}^\prime_0,\hat{\mathbf{y}}^\prime_0)$ as:
$\hat{\mathbf{e}}^E_i=[-1 \quad 0]\T_{(\hat{\mathbf{x}}^\prime_0,\hat{\mathbf{y}}^\prime_0)}$ (H mode) or $[0 \quad 1]\T_{(\hat{\mathbf{x}}^\prime_0,\hat{\mathbf{y}}^\prime_0)}$ (V mode). The antenna coordinate system  $(\hat{\mathbf{x}}^\prime,\hat{\mathbf{y}}^\prime,\hat{\mathbf{z}}^\prime)$ is defined by: $\hat{\mathbf{z}}^\prime=\hat{\mathbf{z}}^\prime_0=-\hat{\mathbf{k}}_i$, $\hat{\mathbf{x}}^\prime$ is collinear to the projection of $\hat{\mathbf{x}}$ on the orthogonal plane $(\hat{\mathbf{x}}^\prime_0,\hat{\mathbf{y}}^\prime_0)$ and $\hat{\mathbf{y}}^\prime=\hat{\mathbf{z}}^\prime \wedge \hat{\mathbf{x}}^\prime$. 
\begin{eqnarray}
 \hat{\mathbf{x}}^\prime&=& \frac{\cos\theta_i \cos\varphi_i \hat{\mathbf{x}}^\prime_0 -\sin\varphi_i\hat{\mathbf{y}}^\prime_0}{\sqrt{\cos^2\theta_i \cos^2\varphi_i+\sin^2\varphi_i}} \\
 \hat{\mathbf{y}}^\prime&=& \frac{ \sin\varphi_i \hat{\mathbf{x}}^\prime_0 + \cos\theta_i \cos\varphi_i \hat{\mathbf{y}}^\prime_0}{\sqrt{\cos^2\theta_i \cos^2\varphi_i+\sin^2\varphi_i}}
\end{eqnarray}

For the H mode, it results in the following expressions for $\hat{\mathbf{e}}^E_i\cdot  \hat{\mathbf{x}}^\prime$ and $\hat{\mathbf{e}}^E_i\cdot  \hat{\mathbf{y}}^\prime$ (and so $\hat{\mathbf{e}}^R_i$):
\begin{eqnarray}
\mathbf{e}^I \cdot  \hat{\mathbf{x}}^\prime &=& \frac{-\cos\theta_i \cos\varphi_i}{\sqrt{\cos^2\theta_i \cos^2\varphi_i+\sin^2\varphi_i}} \\
\mathbf{e}^I \cdot  \hat{\mathbf{y}}^\prime &=& \frac{-\sin\varphi_i}{\sqrt{\cos^2\theta_i \cos^2\varphi_i+\sin^2\varphi_i}}
\end{eqnarray}

Let us sum up the complex scattering coefficient in this concentric polarization diverse acquisition. For the $i$-th polarization acquisition, the complex coefficient $s^\star_n(i)$ is given  by:
\begin{enumerate}
\item \textbf{HH acquisition mode}
\begin{eqnarray} \label{exprHH}
\quad s^\star_n(i) &\approx& [\hat{\mathbf{e}}^R_i\cdot \hat{\mathbf{x}}^\prime \quad \hat{\mathbf{e}}^R_i\cdot \hat{\mathbf{y}}^\prime] \cdot \mathbf{S}_n \cdot
\left[
\begin{array}{c} 
\hat{\mathbf{e}}^E_i\cdot \hat{\mathbf{x}}^\prime \\ 
\hat{\mathbf{e}}^E_i\cdot \hat{\mathbf{y}}^\prime 
\end{array}
\right] \nonumber\\
 &\approx&\frac{1}{\mathcal{K}_i}[\cos^2\theta_i \cos^2\varphi_i\cdot s_n^{\textrm{xx}}+\sin^2\varphi_i\cdot s_n^{\textrm{yy}} \nonumber\\
&& + \cos\theta_i \sin2\varphi_i\cdot s_n^{\textrm{xy}}]
\end{eqnarray} 
\item \textbf{VV acquisition mode}
\begin{eqnarray} \label{exprVV}
 s^\star_n(i) &\approx&\frac{1}{\mathcal{K}_i}[\sin^2\varphi_i\cdot s_n^{\textrm{xx}}+\cos^2\theta_i\cos^2\varphi_i\cdot s_n^{\textrm{yy}} \nonumber\\
&& - \cos\theta_i \sin2\varphi_i\cdot s_n^{\textrm{xy}}]
\end{eqnarray} 
\item \textbf{HV acquisition mode} 
\begin{eqnarray} \label{exprHV}
 s^\star_n(i) &\approx&\frac{1}{\mathcal{K}_i}[\cos\theta\frac{\sin2\varphi}{2}\cdot s_n^{\textrm{xx}}-\cos\theta\frac{\sin2\varphi}{2}\cdot s_n^{\textrm{yy}}\nonumber\\
               & & -(\cos^2\theta\cos^2\varphi-\sin^2\varphi)\cdot s_n^{\textrm{xy}}]
\end{eqnarray} 
\end{enumerate}
with: $\mathcal{K}_i=\cos^2\theta_i \cos^2\varphi_i+\sin^2\varphi_i$.

Afterwards, we consider the whole measurements, i.e. the $M$ observed complex scattering coefficients, in the associated acquisition conditions defined by the successive frequencies $f_i$, azimuth $\theta_i$ and roll $\varphi_i$ for $i \in 1,\cdots,M$. They are stacked in the following complex vector $\bm{\mathcal{S}}$:
\begin{equation}
\bm{\mathcal{S}}=\cro{\mathcal{S}_1 \quad \mathcal{S}_2 \quad \cdots \quad \mathcal{S}_M}\T
\end{equation} 

\subsection{Derived direct matrix model}
We now derive the observation model (forward model) from the previous extended MS model. Let us stress again that the "small bandwidth small angle" is assumed: the scattering matrices are supposed to be nearly constant for each elementary scatterer. The unknown  is made up of the scattering matrices associated to the $N$ discretized scatter points $\mathbf{r}_n$ ($n=1,\cdots,N$) of a 3D grid, the coordinates of which are $(x_n,y_n,z_n)$ in the target coordinate system $(O\hat{\mathbf{x}}\hat{\mathbf{y}}\hat{\mathbf{z}})$. The scattering matrices can be separated in three 3D maps $\bm{s}_{\textrm{xx}}$, $\bm{s}_{\textrm{yy}}$ et $\bm{s}_{\textrm{xy}}$ (in the target reference frame), respectively for $\textrm{HH}$, $\textrm{VV}$ and $\textrm{HV}$ antenna polarization mode (defined at the initial reference rotation, for $\theta=0^\circ$ and $\varphi=0^\circ$). In vector form, they can be expressed respectively by: 
$\bm{s}_{\textrm{xx}} = \cro{ s_1^{\textrm{xx}} \quad s_2^{\textrm{xx}} \quad \cdots \quad s_N^{\textrm{xx}} }\T$, $\bm{s}_{\textrm{yy}} = \cro{ s_1^{\textrm{yy}} \quad s_2^{\textrm{yy}} \quad \cdots \quad s_N^{\textrm{yy}} }\T$ and $\bm{s}_{\textrm{xy}} = \cro{ s_1^{\textrm{xy}} \quad s_2^{\textrm{xy}} \quad \cdots \quad s_N^{\textrm{xy}} }\T$.
From (\ref{Eq:MS_ext}), the complex scattering coefficients $\mathcal{S}_i$ are clearly related to the Fourier transform of the $s^\star_n$ coefficients. Furthermore, it is obvious from (\ref{exprHH}) and (\ref{exprVV}) that the scattering coefficients are linear combinations of components $\bm{s}_k$ ($k \in \{\textrm{xx},\textrm{yy},\textrm{xy}\}$). Therefore, the direct model (\ref{Eq:MS_ext}) can be rewritten as (with noise vector $\bm{ \varepsilon}$):
\begin{eqnarray} 
\bm{\mathcal{S}} &=& \sum_{k \in \{xx,yy,xy\}} \Tb\Wb_k\Fb\bm{s}_k + \bm{ \varepsilon} 
\end{eqnarray} 
where $\Fb$ is a DFT (Discrete Fourier Transform), that can be computed by FFT (Fast Fourier Transform) algorithm. Finally, $\Tb$ is the truncation matrix related to the position of the observed points in Fourier domain. The three matrices $\Wb_k$ are diagonal weight matrices that affect Fourier coefficients ($k \in \{\textrm{xx},\textrm{yy},\textrm{xy}\}$): 
\begin{equation} 
\Wb_k=\mathbf{diag}(\mathrm{w}_k(1),\mathrm{w}_k(2),\cdots,\mathrm{w}_k(M))
\end{equation} 

 Depending on the polarization acquisition mode (HH, VV or HV), the weights $\Wb_k$ are given respectively by (\ref{exprHH}), (\ref{exprVV})  and (\ref{exprHV}). They define the elements or combination of  elements of the $N$ MS scattering matrices (i.e. $s_n^{\textrm{xx}}$, $s_n^{\textrm{yy}}$ or $s_n^{\textrm{xy}}$, for $n=1,\cdots,N$) which determine the scattering coefficient observation $\mathcal{S}_i$. When the acquisition mode is $\textrm{HH}$, the weights are given by (for $i=1,\cdots,M$): 
\begin{equation} \label{Eq:Weights}
\left\{
\begin{array}{l} 
\mathrm{w}_{\textrm{xx}}(i)=\cos^2\theta_i\cos^2\varphi_i/\mathcal{K}_i\\
\mathrm{w}_{\textrm{yy}}(i)=\sin^2\varphi_i/\mathcal{K}_i\\
\mathrm{w}_{\textrm{xy}}(i)=\cos\theta_i \sin2\varphi_i /\mathcal{K}_i
\end{array} 
\right.
\end{equation} 

\begin{figure}[ht!] \centering 
\includegraphics*[width=0.5\textwidth] {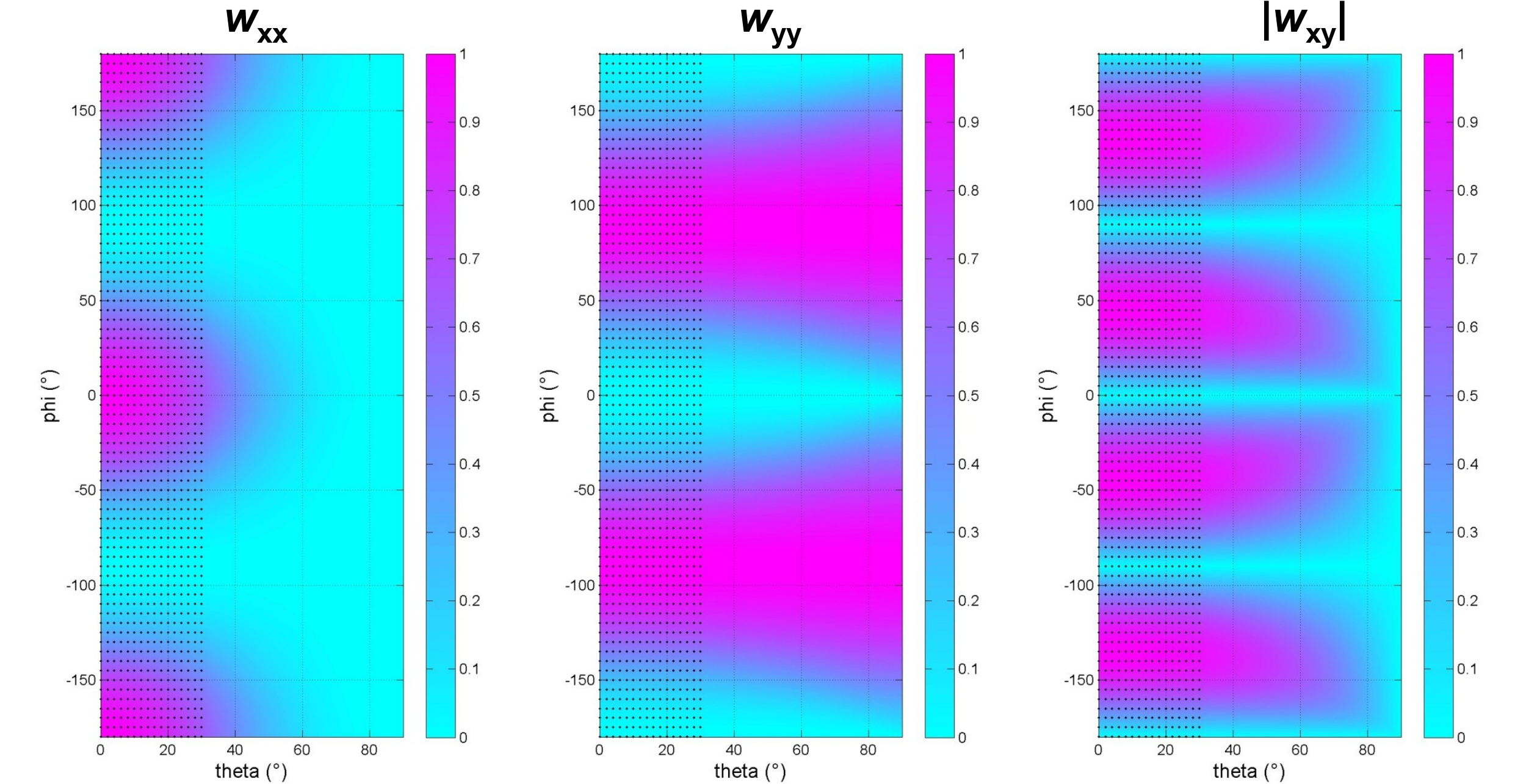}
\caption{Weigths $\mathrm{w}_{\textrm{xx}}$, $\mathrm{w}_{\textrm{yy}}$ and $|\mathrm{w}_{\textrm{xy}}|$ in function of $\theta$ and $\varphi$, with typical acquisition angles (black points).} \label{weights}
\end{figure}

They are shown in Fig.~\ref{weights}, for various acquisition angles $\theta$ and $\varphi$. Typical acquisition angles ($\theta \le 30^\circ$ and $0^\circ \le \varphi \le 360^\circ$) are represented by black points. Notice that the weights $\mathrm{w}_{\textrm{xx}}$, $\mathrm{w}_{\textrm{yy}}$ and $\mathrm{w}_{\textrm{xy}}$ depend mainly on the value of $\varphi$. For $\varphi$ is close to $0^\circ$, only the $s_n^{\textrm{xx}}$ terms matter. Conversely, when  $\varphi$ close to $\pm90^\circ$, only the $s_n^{\textrm{yy}}$ matter and for that reason are observable. Notice that it is true as long as $\theta \le 40^\circ$. For larger $\theta$, the scattering terms $s_n^{\textrm{xx}}$ no longer affect the coherent superposition echo.  Anyway, the  "small bandwidth small angle" assumption of constant $\mathbf{S}_n$ will be violated for common complex targets.

Finally, merging the 3 maps in an unique column vector  $\bm{s} = \cro{\bm{s}_{\textrm{xx}}; \bm{s}_{\textrm{yy}}; \bm{s}_{\textrm{xy}}}$, the final direct model reads: 
\begin{equation} \label{DM_mat}
\bm{\mathcal{S}}  = \Ab \bm{s} + \bm{ \varepsilon} 
\end{equation} 
where $\Ab$ is a large block matrix of size $M\times 3N$:
\beq \label{Eq:MatriceA}
\Ab = \left[
\begin{array}{c|c|c}  \Tb\Wb_{\textrm{xx}}\Fb & \Tb\Wb_{\textrm{yy}}\Fb & \Tb\Wb_{\textrm{xy}}\Fb \end{array} 
\right]
\eeq
The observation matrix $\Ab$ defines the deterministic part of the relation between the observations and the unknown state vector made of the three vectorized 3D map. It must be noticed that its dimensions can be huge, \eg $N\approx 500 000$ and $M=15\cdot10^6$. And yet, the model (\ref{DM_mat}) is still linear (linear transforms and additive noise). Moreover, it relies on simple and fast transform (FFT, weight, truncation). Again, remark that a close parallel can be made with polarimetric SAR \cite{Lee,zebker1991imaging}, with a linear and additive noise model and a polarimetric measurement matrix which, similarly to $\Ab$, is determined by the successive polarizations at emission and reception.

\subsection{Regularized inversion and Minimum Norm Least Squares} 

%
Regarding the inverse problem, a common approach relies on a discrepancy between measurements and model outputs. A standard discrepancy is based on a quadratic norm and yields the so called Least Squares (LS) criterion: 
\beq \label{Eq:CritMC}
\Jc\MC(\bm{s}) = \norm{\bm{\mathcal{S}} -\Ab \bm{s}}^2 \,.
\eeq
A potential solution minimizes~$\Jc\MC$, \ie it is the one that best fits the data: 
\beq \label{Eq:MCArgMin}
\wh{\bm{s}}\MC = \argmin_{\bm{s}\in\eC^N} \Jc\MC(\bm{s})  \,.
\eeq
The minimizer $\wh{\bm{s}}\MC$ can be found by setting the gradient of $ \Jc\MC(\bm{s})$ to zero and it leads to a linear system: 
\beq \label{Eq:EqNormaleMC}
\Ab^\dag \Ab \bm{s} = \Ab^\dag \bm{\mathcal{S}}  \,.
\eeq
Unfortunately, it cannot be solved since $\Ab^\dag \Ab$ is rank-deficient (the number of unknowns is larger than the number of measurements) and, as a consequence, an infinity of objects minimizes (and nullifies) the LS criterion. In other words, an infinity of backscattered maps is exactly consistent with the data. 
Among these solutions, a possible simple approach selects the one with minimum norm:
\beq \label{Eq:PbIOptimMCNM}
\wh{\bm{s}}\MCNM = 
\bca
\displaystyle \argmin_{\bm{s}\in\eC^N} \norm{\bm{s}}^2 \cr 
\ST~ \bm{\mathcal{S}} -\Ab\bm{s}=0
\eca
\eeq
\ie the Minimum Norm Least Squares (MNLS) solution. 
There is a very large literature on various methods to deal with this kind of situations and build other solutions. The interested reader can, for instance, refer to books on inverse problems \cite{Giovannelli15,Bertero02,Chalmond03,Kaipio05,Aster05}. 
Coming back to our simple MNLS solution, it can be shown that 
\beq \label{Eq:OptimMCNM}
\wh{\bm{s}}\MCNM  =  \Ab^\dag (\Ab\Ab^\dag)\pmu \bm{\mathcal{S}} \,.
\eeq
A possible proof is based on Lagrange multipliers and it is given in Appendix. 
The estimated map $\wh{\bm{s}}\MCNM$ is a linear transform of the data, nevertheless the matrix $\Ab\Ab^\dag$ is huge and its inverse cannot be computed. Fortunately, it can be inverted analytically: given the specificity of~(\ref{Eq:MatriceA}), we have
\beqx
\Ab\Ab^\dag = \Tb\, (\Wb_{\textrm{xx}}^2 + \Wb_{\textrm{yy}}^2 +\Wb_{\textrm{xy}}^2) \,\Tb\T 
\eeqx 
that is a diagonal matrix and hence easily invertible. 
Finally, relation~(\ref{Eq:OptimMCNM}) becomes
\beqx
\wh{\bm{s}}\MCNM = 
\left[
\begin{array}{c}  \Fb^\dag\Wb_{\textrm{xx}}\Tb\T \\ \Fb^\dag\Wb_{\textrm{yy}}\Tb\T \\ \Fb^\dag\Wb_{\textrm{xy}}\Tb\T   \end{array} 
\right] 
\cro{\Tb\, (\Wb_{\textrm{xx}}^2 + \Wb_{\textrm{yy}}^2 +\Wb_{\textrm{xy}}^2) \,\Tb\T}\pmu \bm{\mathcal{S}} 
\eeqx 
that jointly determines the three 3D backscattered maps associated with the $\textrm{xx}$, $\textrm{yy}$ and $\textrm{xy}$ polarizations. In addition, each individual map $\wh{\bm{s}}^k\MCNM$ is obtained separately by: 
\beqx \label{Eq:MCPTriFourier}
\wh{\bm{s}}^k\MCNM = \Fb^\dag \Wb_k \Tb\T \cro{ \Tb\, (\Wb_{\textrm{xx}}^2 + \Wb_{\textrm{yy}}^2 +\Wb_{\textrm{xy}}^2)  \,\Tb\T }\pmu \bm{\mathcal{S}} 
\eeqx
from the data set $\bm{\mathcal{S}}$. It can be seen that they are obtained by IFFT of the zero-padded weighted measurements. The involved weights are:
\beq \label{Eq:TransfWeight}
\pi_k(m)= 
\frac{w_k(m)}{ w_{\textrm{xx}}(m)^2 + w_{\textrm{yy}}(m)^2 +w_{\textrm{xy}}(m)^2} 
\eeq
based on direct weight~(\ref{Eq:Weights}), that depends on the polarization mode and the acquisition angles (azimuth and roll). 

\begin{figure}[ht!]
\centering \includegraphics[width=0.5\textwidth,clip=true,viewport=2cm 1.5cm 25cm 18cm]{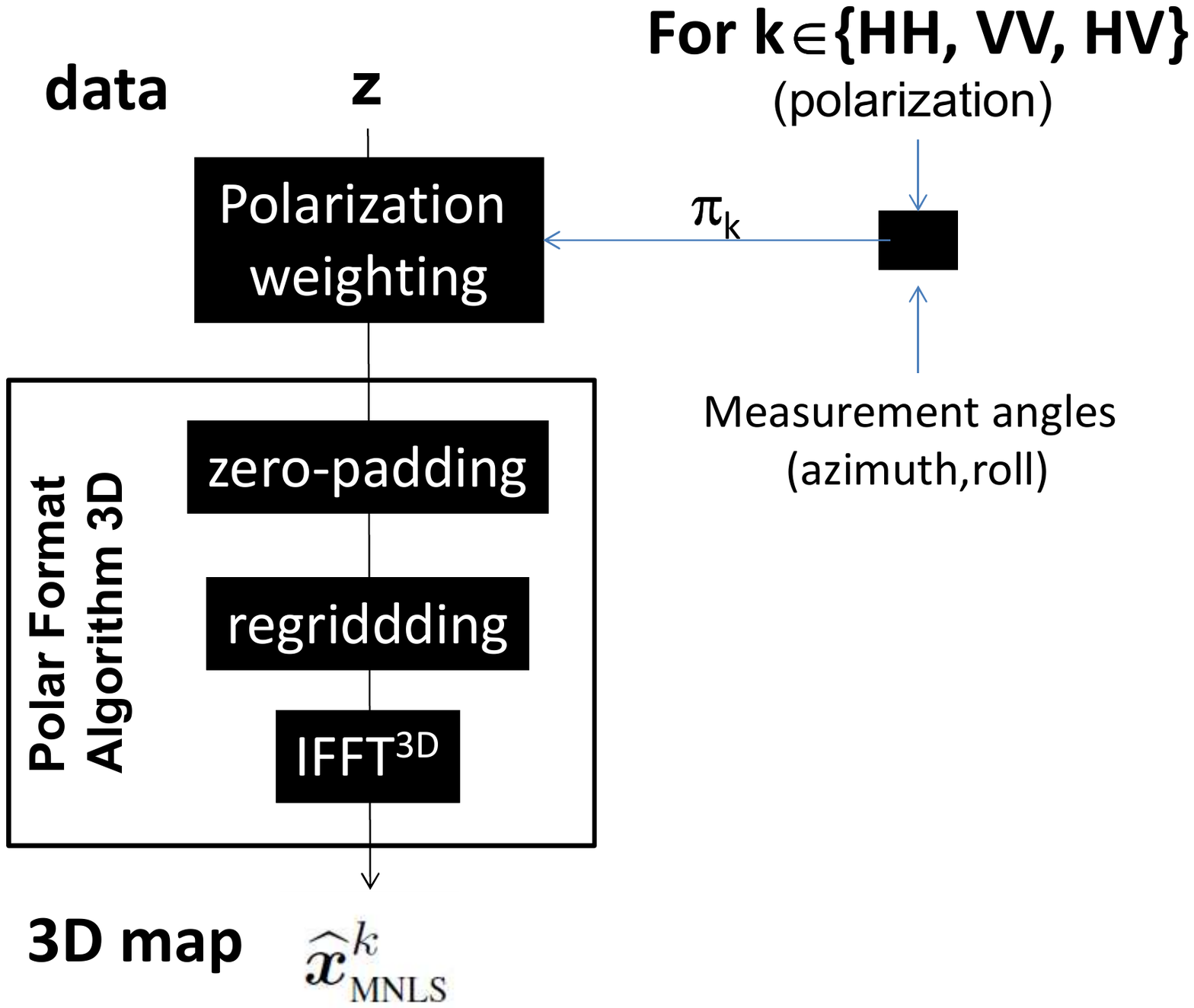} \caption{Fast regularized inversion} \label{Fig:Algo}
\end{figure}

The resulting fast 3D radar imaging algorithm is composed of two steps as described in Fig.~\ref{Fig:Algo}. The first step consists in processing the data by applying the weights of Eq.~(\ref{Eq:TransfWeight}), for each polarization $\textrm{xx}$, $\textrm{yy}$ and $\textrm{xy}$, and for each associated weights. Afterwards, in the second step, three 3D backscattered maps, associated with $\textrm{xx}$, $\textrm{yy}$ and $\textrm{xy}$ (in the target reference frame), are separately reconstructed by a 3D PFA method \cite{soumekh1999synthetic,cheney2009fundamentals}. It includes regridding and inverse 3D IFFT steps. It is finally a very efficiently implementation, based on weight, zero-padding and IFFT.

\section{Application} \label{Section_Application}

Indoor 3D spherical near field facilities are dedicated to various tasks. Beyond scattering analysis with 2D or 3D ISAR imaging, they are encountered for RCS measurement, microwave absorber measurement, nondestructive testing, material characterization, antenna measurement and diagnosis, research in near-field measurement techniques, etc. These facilities are generally based on a mobile arch that provides various illumination viewpoints. Most of them are dedicated to antenna measurements; let us mention for example the commercial systems "G-DualScan" and "700S-90", respectively developed by Satimo and NSI. Others are specialized in RCS measurement. See, for instance, the system of the French research institute Fresnel at the Microwave Common Resources Centre (MCRC), equipped with five positioners that allows spherical diffraction patterns \cite{eyraud2008validation}, as well as the bistatic anechoic chamber BIANCHA \cite{escot2010indoor} of the Italian institute INTA. It provides a bistatic, spherical field scanner based on a dual-axis azimuth turntable and  two  elevated scanning arms. Related to these 3D spherical facilities, our 3D spherical RCS set-up, designed and developed at CEA, is specifically dedicated to RCS analysis of small targets of lower RCS. After a short description of the facility and its microwave instrumentation, we present 3D radar imaging results on various mock-ups. 

\subsection{The 3D RCS CEA facility}
\begin{figure}[!ht]
\centering \includegraphics[width=0.3\textwidth, clip=true]{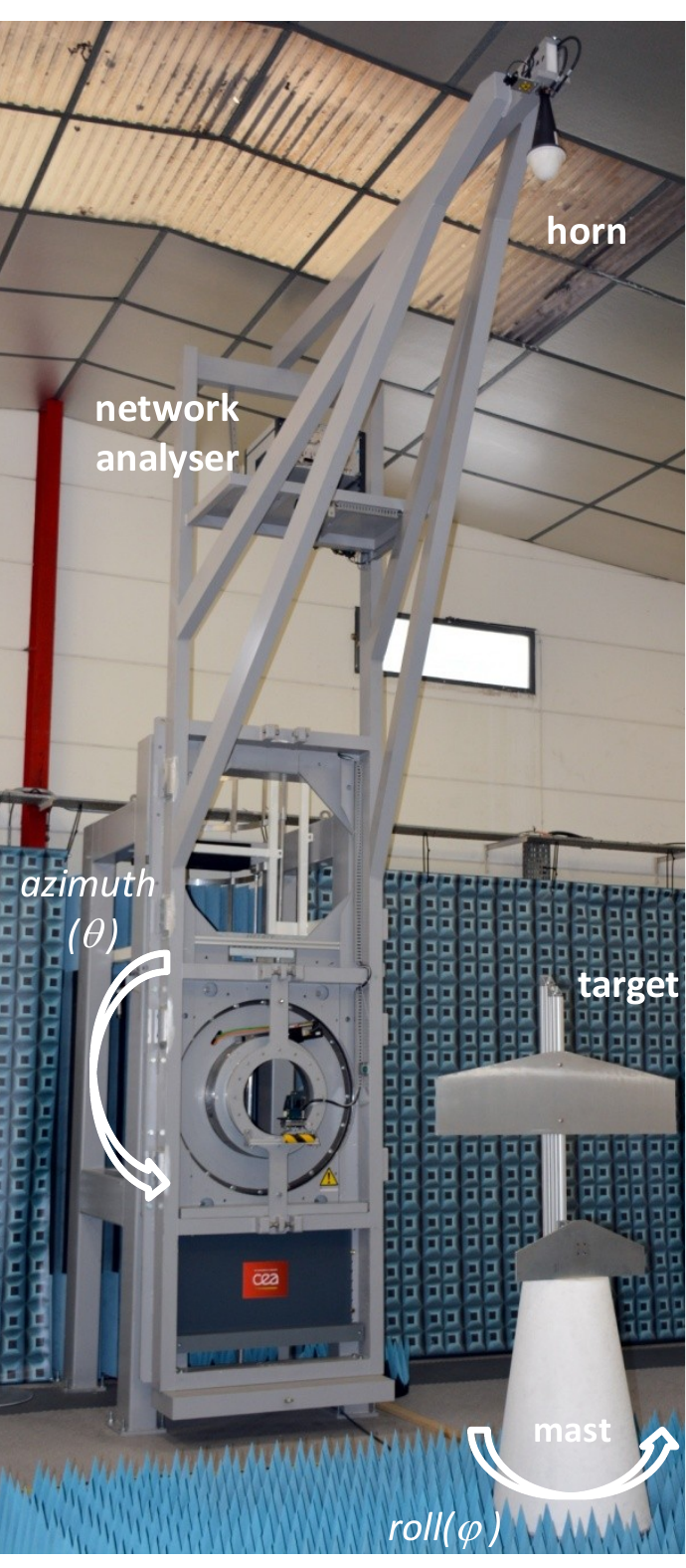} \caption{Near-field 3D RCS measurement facility} \label{Arch3D}
\end{figure}

The concentric polarization-diverse measurements of previous section \ref{sec_polardiv} are achieved by the experimental layout of Fig.~\ref{Arch3D}. Dedicated to the RCS characterization of small targets (typically $\le 1$ ft), it is composed of a 4 meters radius motorized rotating arch (around horizontal axis) holding the measurement antennas while the target is located on a polystyrene mast mounted on a rotating positioning system (around vertical axis). The combination of the two rotation capabilities leads to a full 3D near field monostatic RCS characterization. 

The microwave instrumentation is made of two bipolarization monostatic RF transmitting and receiving antennas that are driven by a fast network analyzer. A phased array antenna is used for the 0.8~-1.8 GHz frequency band. It is optimized in order to reduce spurious signals originated from interactions with the arch metallic structure. For higher frequencies from 2 to 12 GHz, a wideband standard gain horn, equipped with a lens, is preferred (see Fig.~\ref{Arch3D}). The target under test is located vertically on a polystyrene mast below the measurement antenna. The target rotation around vertical axis is determined by the roll angle $\varphi$. The arch rotation around horizontal axis is determined by the azimuth angle $\theta$. A full set of data $\bm{\mathcal{S}}$ is therefore composed of successive frequency, azimuth, and roll sweeps. The arch is built of aluminum, weights about 200 kg, measures approximately 5 m $\times$ 3 m, and is designed to minimize potential mechanical deformations and to hold the measurement antenna over a total travel range of $\pm 100^o$. The $\theta$ rotation is achieved using a direct drive positioner including a brushless motor with a high accuracy encoder directly coupled to the motor for accuracy and repeatability concerns. The $\varphi$ rotation uses another positioner also including a brushless motor with a high accuracy encoder directly coupled to the motor. The polystyrene mast \cite{Massaloux4,Massaloux5} supporting the target under test is located on this rotating positioning system (vertical axis). Let us emphasize the precision of the whole positioning system. The two major errors are the residual radial error and the repositioning (repeatability) error of the arch. The first one is about $\pm 0.001^o$ and is equivalent to a maximum error of $\pm 0.75$ mm on the vertical axis. The second one is about $\pm 0.15$ mm maximum, once again on the vertical axis. Refer to \cite{Massaloux1} for details on the laser tracker characterization.


Concerning the measurement process, it is composed of successive stages. First, a calibration by substitution \cite{Knott_m} is performed: it consists in replacing the test target with a calibration standard whose echo is well known in order to determine the inverse transfer function $\mathcal{K}$ of the entire RCS measurement system. This enables to convert indirectly the received transmitted electric field $T_{out}$ to a complex scattering coefficient value: 
$\mathcal{S}=\mathcal{K}.T_{out}$. Basically, the substitution method establishes a “phase reference” relatively to the rotation center and normalizes out the dispersive frequency response of the RCS measurement system. Note that for large objects, it can be enhanced by a multi-calibration approach \cite{Massaloux3} which overcomes most of near-field effects, such as the power decrease or the EM wave sphericity. The second stage corresponds to background subtraction \cite{Knott_m}. It is commonly introduced when the clutter is too high, compared to the test target and calibration signals. It removes coherently the background echoes. Eventually, an adaptive range filter is applied to eliminate the residual interferences, e.g. interactions with walls and floor that can affect the useful signal.

\subsection{Results}
\begin{figure}[!ht]
\centering \includegraphics[width=0.3\textwidth, clip=true]{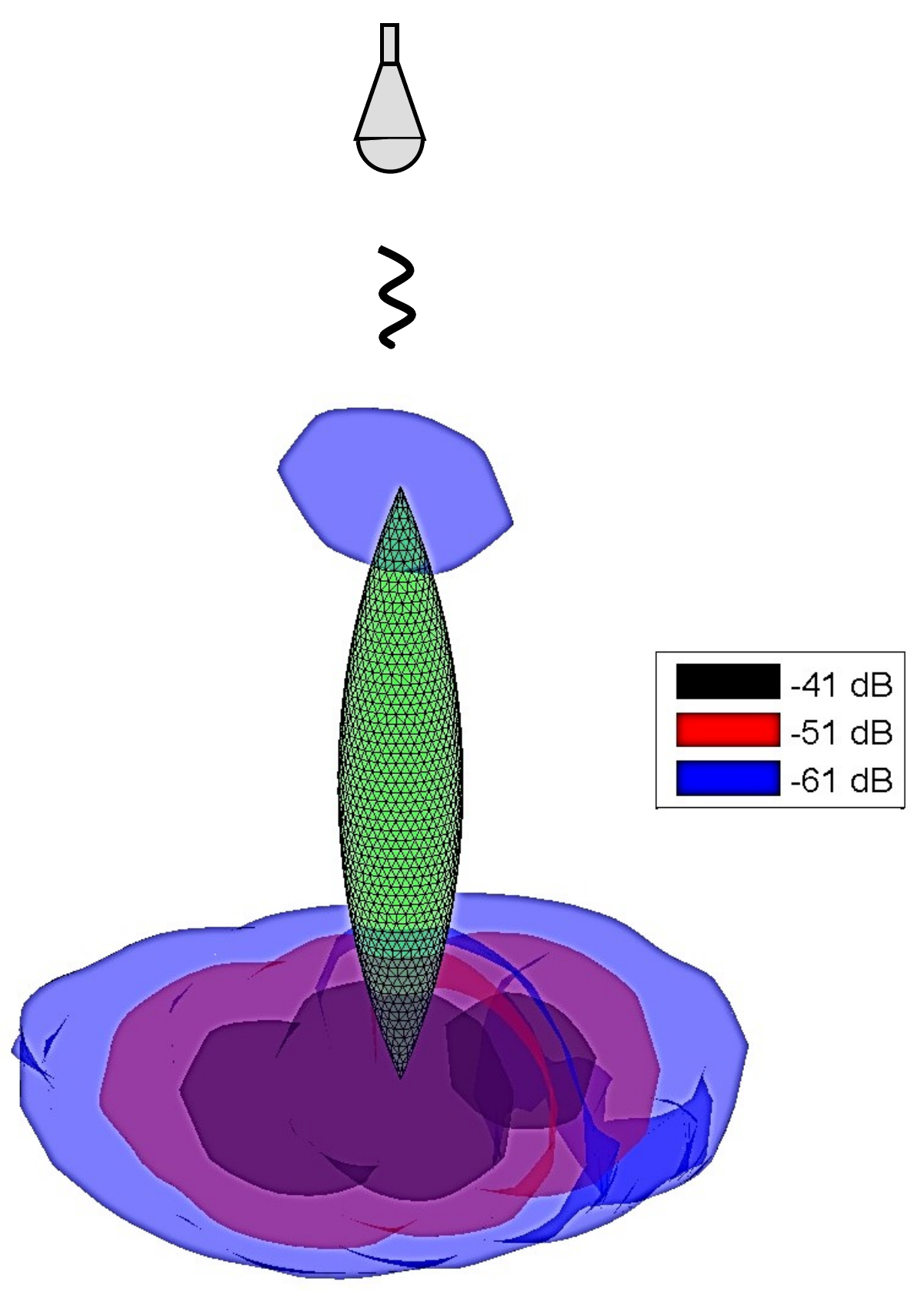} \caption{Ogival shape  3D radar map $\wh{\bm{s}}^\textrm{xx}\MCNM$ [$N=64\times64\times1024,M\approx2\cdot10^5$]} \label{ImagOgival}
\end{figure}

Let us first illustrate our 3D radar imaging method by an application to the following synthetic RCS data. We consider the metallic ogival-shaped object from the EM benchmark \cite{woo1993programmer}, with polarization-diverse measurements at $\theta\in[0^\circ:2^\circ:20^\circ], \varphi\in[0^\circ:5^\circ:360^\circ[$  and $f\in[1 \mbox{ GHz}:10 \mbox{ MHz}:3  \mbox{ GHz}]$. The synthetic data is computed by an efficient parallelized harmonic Maxwell solver\footnote{It combines a volume finite element method and integral equation technique, taking benefit from the axisymmetrical geometry of the shape \cite{giraud2013advanced}}. Fig.~\ref{ImagOgival} shows the 3D scatterer map $\wh{\bm{s}}^\textrm{xx}\MCNM$, corresponding to $\textrm{xx}$ polarization. The 3D map corresponds to the indicated above radar viewpoint, along $O\hat{\mathbf{z}}$. Here, the 3D volume is  represented by several transparent isosurfaces, that are linked to different scatterer levels. In this basic situation, the main EM scatterers are located: the diffraction on the above tip and, at a higher level, the diffraction on the below tip. It comes with sidelobes that derives from the limited acquisition sampling. 

Next, it is shown how the 3D radar imaging approach can support RCS analysis in more complex situations from real polarization-diverse measurement data. Various targets have been characterized in order to evaluate the 3D radar imaging method: a metallic cone with patches, a metallic arrow and a glider. Each of them enables to test the capacity to deal with  various target-wave interaction phenomena: localized scatterers  with the metallic cone, multiple diffractions as well as different polarization responses with the metallic arrow and near field effects due to the large wingspread of the glider mock-up (regarding to the wave number). All the scattering measurements have been acquired for a $360^\circ$ roll sweep ($\varphi$) and a $\pm20^\circ$ azimuth sweep ($\theta$), from  $8.2$ to $12.4$ GHz frequency. 

\subsubsection{Localized scatterers}
\begin{figure}[!ht]
\centering \includegraphics[width=0.5\textwidth, clip=true]{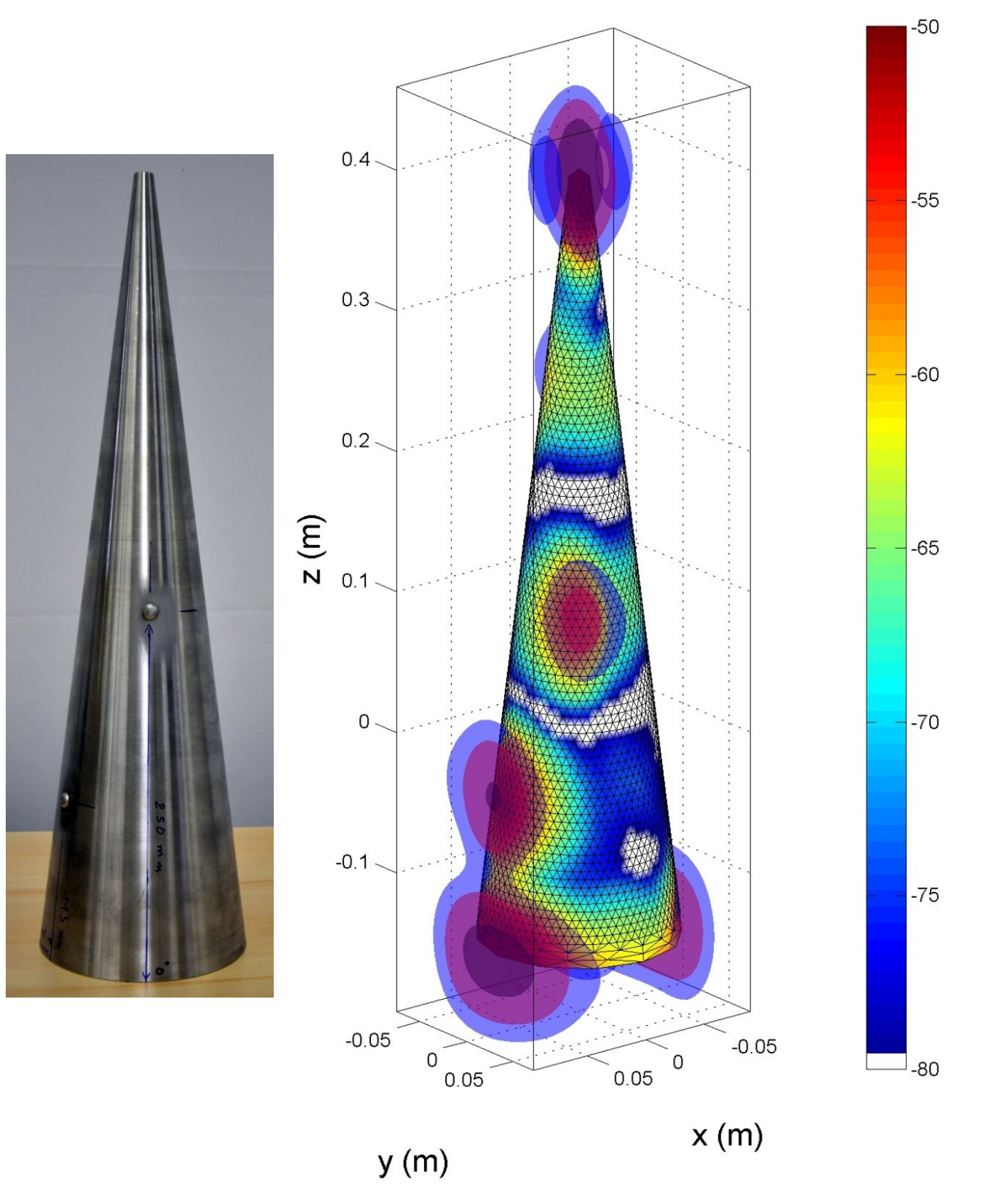} \caption{Metallic cone with 3 metallic patches (left) - 3D radar map $\wh{\bm{s}}^\textrm{xx}\MCNM$ [$N=256\times256\times512,M\approx 10^5$] (right)} \label{coneIEEE}
\end{figure}

Fig.~\ref{coneIEEE}  shows the 3D radar map $\wh{\bm{s}}^\textrm{xx}\MCNM$ in polarization $\textrm{xx}$ of a metallic cone (height: 60cm) where 3 metallic patches (1cm$\times$4cm) have been glued to points $z=250$ mm ($-45^\circ$ roll), $z=400$ mm ($135^\circ$ roll) and $z=112$ mm ($270^\circ$ roll). For convenience, the 3D volume is  represented by several isosurfaces, superimposed upon a volume section relatively to the target shape. The main scatterers are perfectly located: the tip, the diffraction of the rear edge and each metallic patch. Notice that, since the frequency band is here larger than for the previous ogival-shaped object, the resolution is improved along $O\hat{\mathbf{z}}$.  The scatterers corresponding to the 3 metallic patches are correctly located. 

\begin{figure}[!ht]
\centering \includegraphics[width=0.4\textwidth, clip=true]{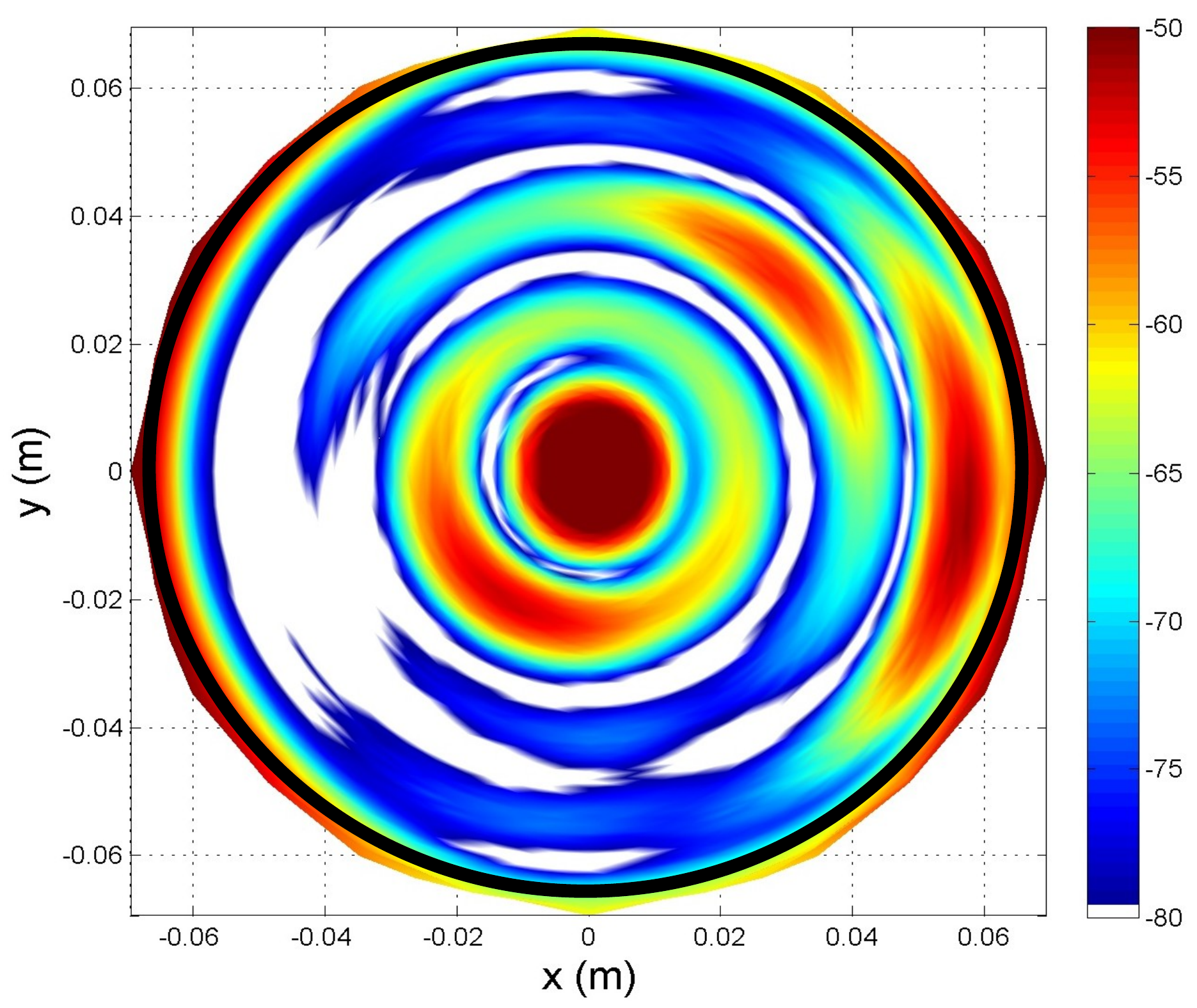} \caption{Above view of the 3D radar map $\wh{\bm{s}}^\textrm{xx}\MCNM$ (shape section) [$N=256\times256\times512,M\approx 10^5$]} \label{VD_coneIEEE}
\end{figure}

It is confirmed in the above view of Fig.~\ref{VD_coneIEEE}, where the image can be compared to the photo, as well as in the exploded image of Fig.~\ref{coneIEEE}. There, the true patch locations are represented by black pellets. Notice that although the creeping waves are not shown in Fig.~\ref{coneIEEE}, they would appear, under the cone, in other representations.

\begin{figure}[!ht]
\centering \includegraphics[width=0.4\textwidth, clip=true]{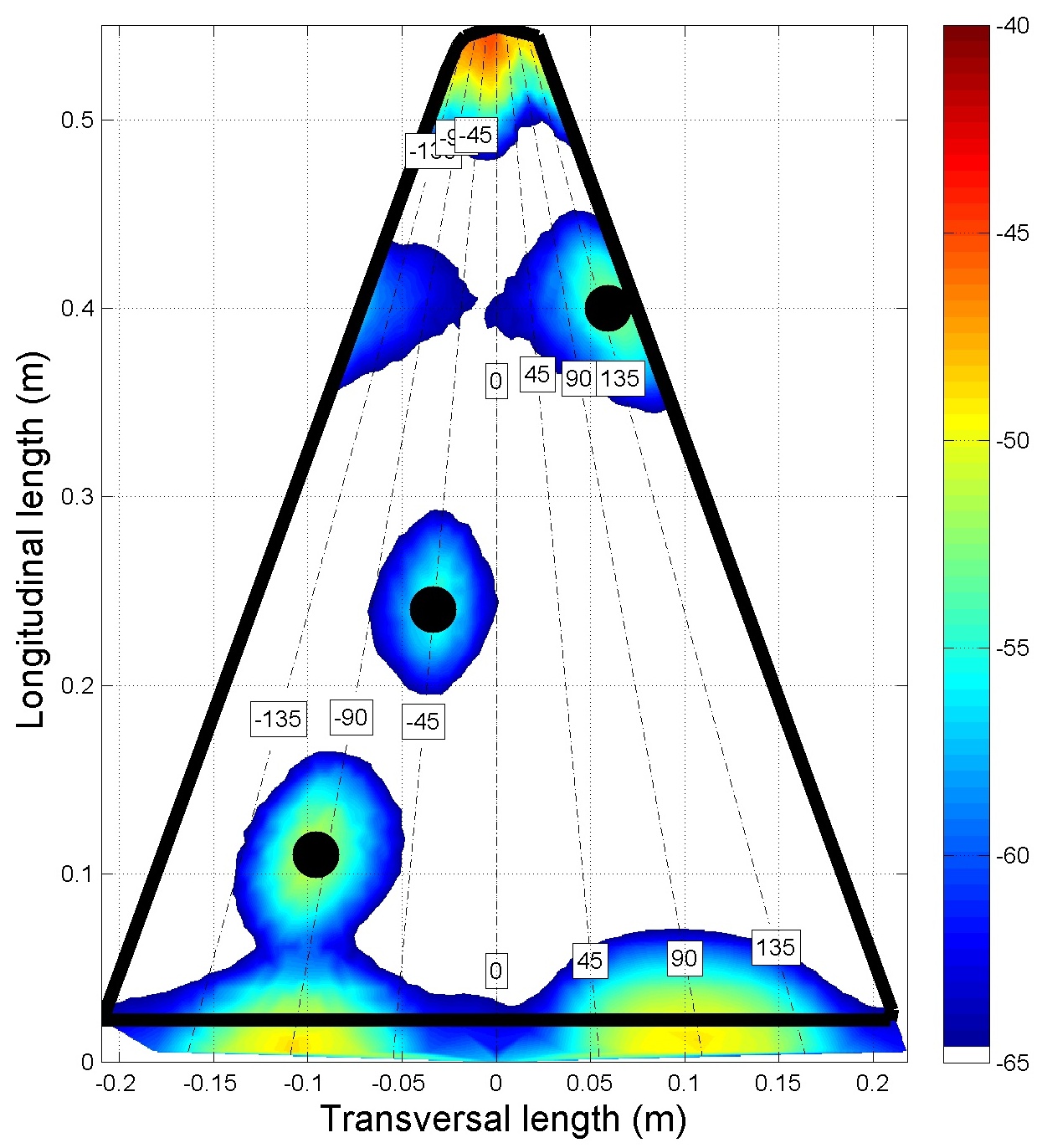} \caption{Exploded image of the 3D radar map $\wh{\bm{s}}^\textrm{xx}\MCNM$ [$N=256\times256\times512,M\approx 10^5$]} \label{coneIEEE_eclate}
\end{figure}

\begin{figure}[!ht]
\centering \includegraphics[width=0.25\textwidth, clip=true]{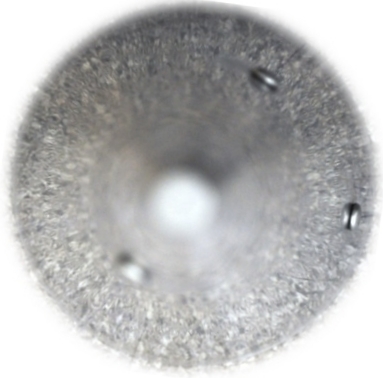} \caption{Above view of the metallic cone with 3 metallic patches} \label{VD_coneIEEEr}
\end{figure}

\begin{figure}[!ht]
\centering \includegraphics[width=0.5\textwidth, clip=true]{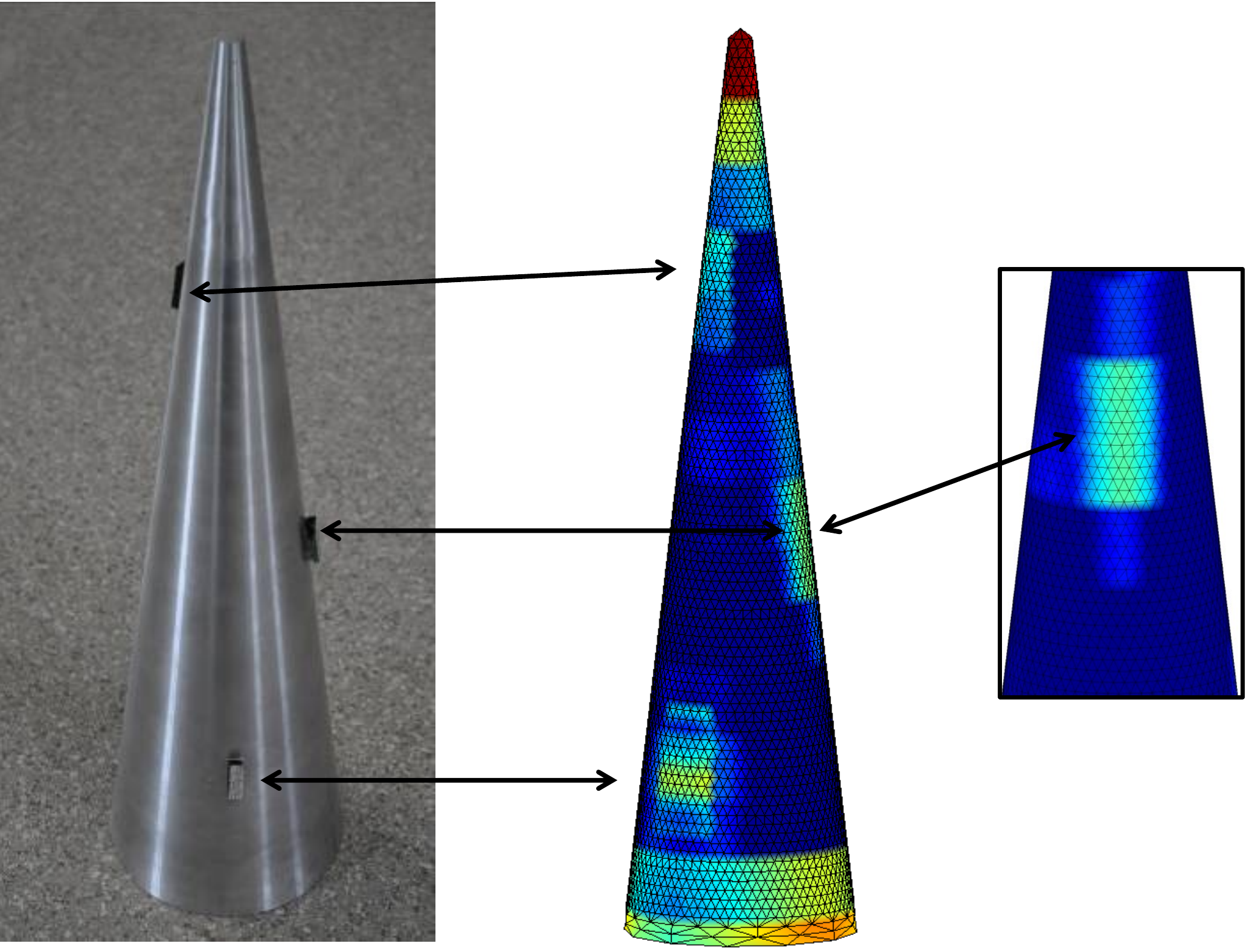} \caption{Metallic cone with 3 RAM patches  - 3D radar map $\wh{\bm{s}}^\textrm{xx}\MCNM$ (shape section) [$N=256\times256\times512,M\approx 10^5$]} \label{RAMCone}
\end{figure}

Next, the metallic patches are replaced by  3 RAM (Radar Absorbing Material) rectangular patches (1cm$\times$4cm) which have been glued to the same locations. Results are presented in Fig.~\ref{RAMCone}. If the scatterers associated to the 3 RAM patches are correctly located, they are now far more diffuse; it is then more difficult to evaluate the location accuracy of the 3D imaging method. This phenomenon can be explained by the fact that RAM has a spread frequency response, which does not allow to obtain a very localized scatterer. 

 \subsubsection{Multiple diffraction \& polarization dependence} The second target is a metallic arrow of 472 mm long and 150 mm large (it belongs to Airbus Group Innovation, see \cite{castelli2014national} for details). It is represented in Fig.~\ref{RAMCone}. With such a target, diffraction edges are known to differ according to the polarization of the incident wave. Moreover, multiple diffractions are likely to appear between the back of the arrow and its base. 
 
 

\begin{figure}[!ht]
\begin{center}
\subfigure{\includegraphics[width=0.23\textwidth, clip=true]{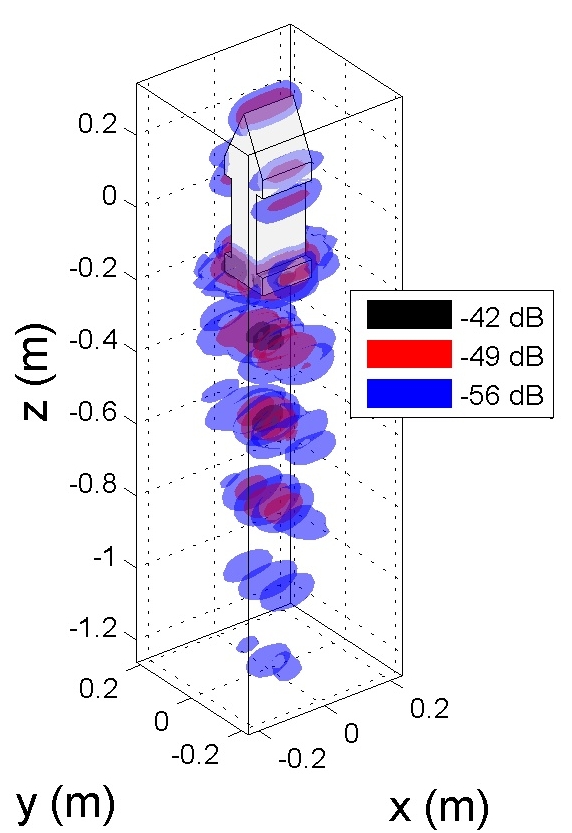}}
\subfigure{\includegraphics[width=0.23\textwidth, clip=true]{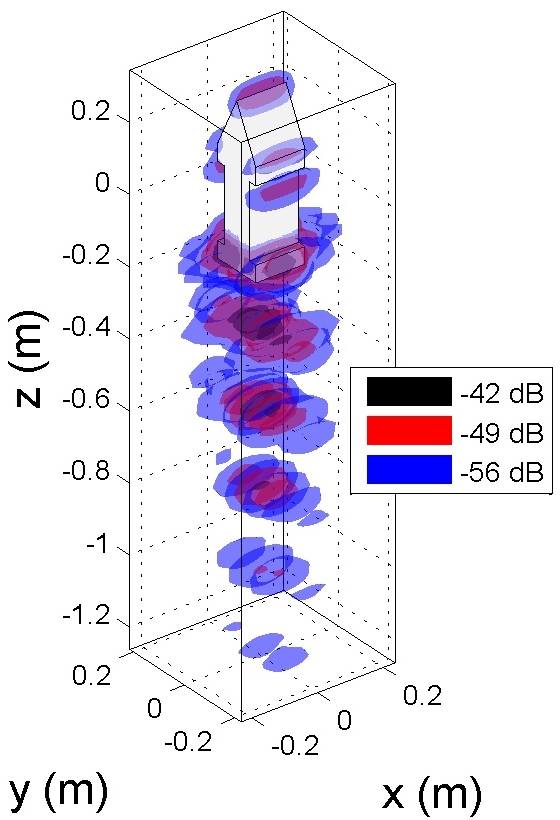}}
\end{center}
\caption{3D radar image  ($\wh{\bm{s}}^\textrm{xx}\MCNM$) of the arrow, from real data (left) and simulated data (right) [$N=40\times40\times1146,M\approx 4\cdot10^6$]}\label{fleche3D_yy}
\end{figure}

\begin{figure}[!ht]
\begin{center}
\subfigure{\includegraphics[width=0.23\textwidth, clip=true]{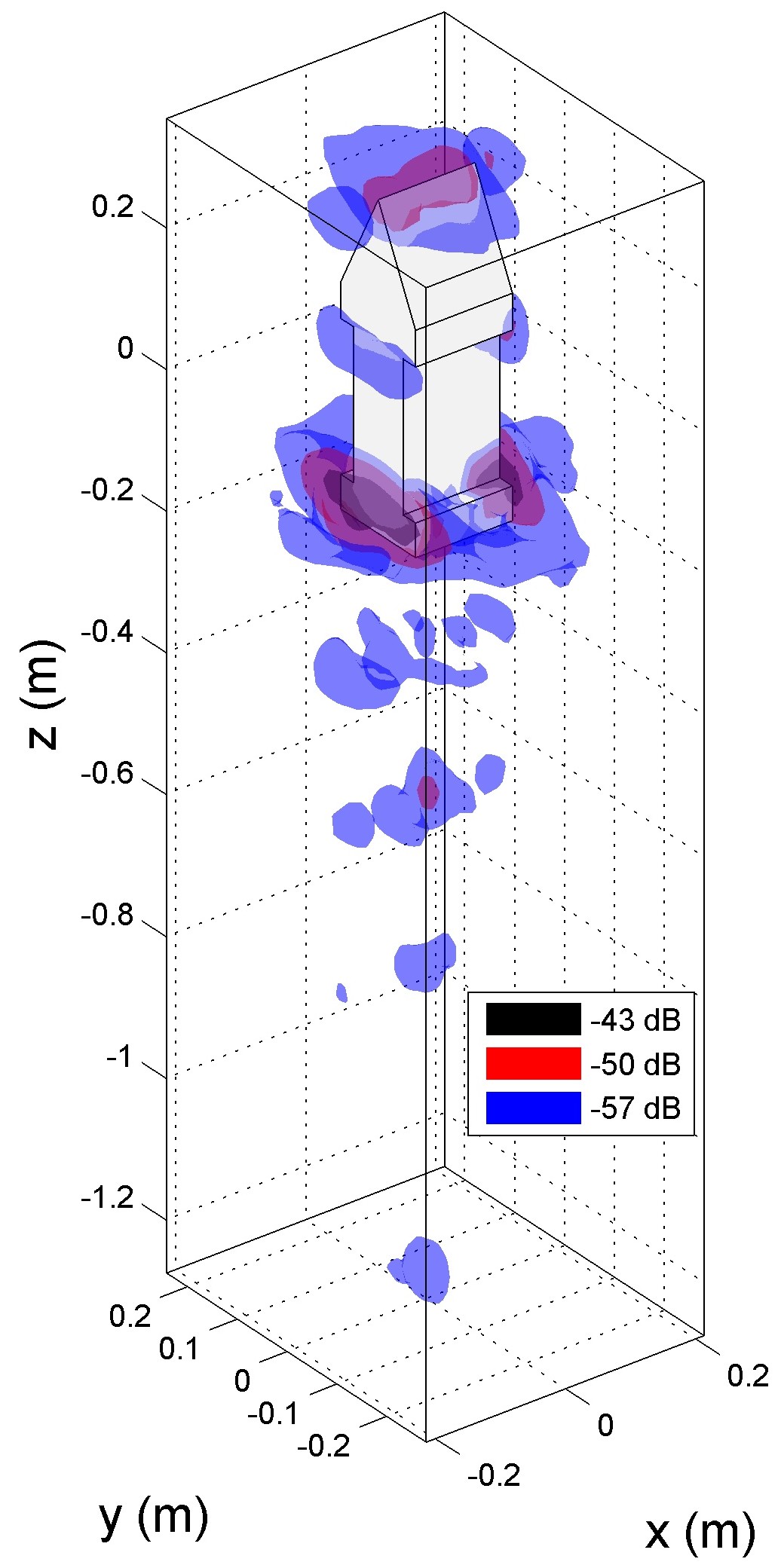}}
\subfigure{\includegraphics[width=0.24\textwidth, clip=true]{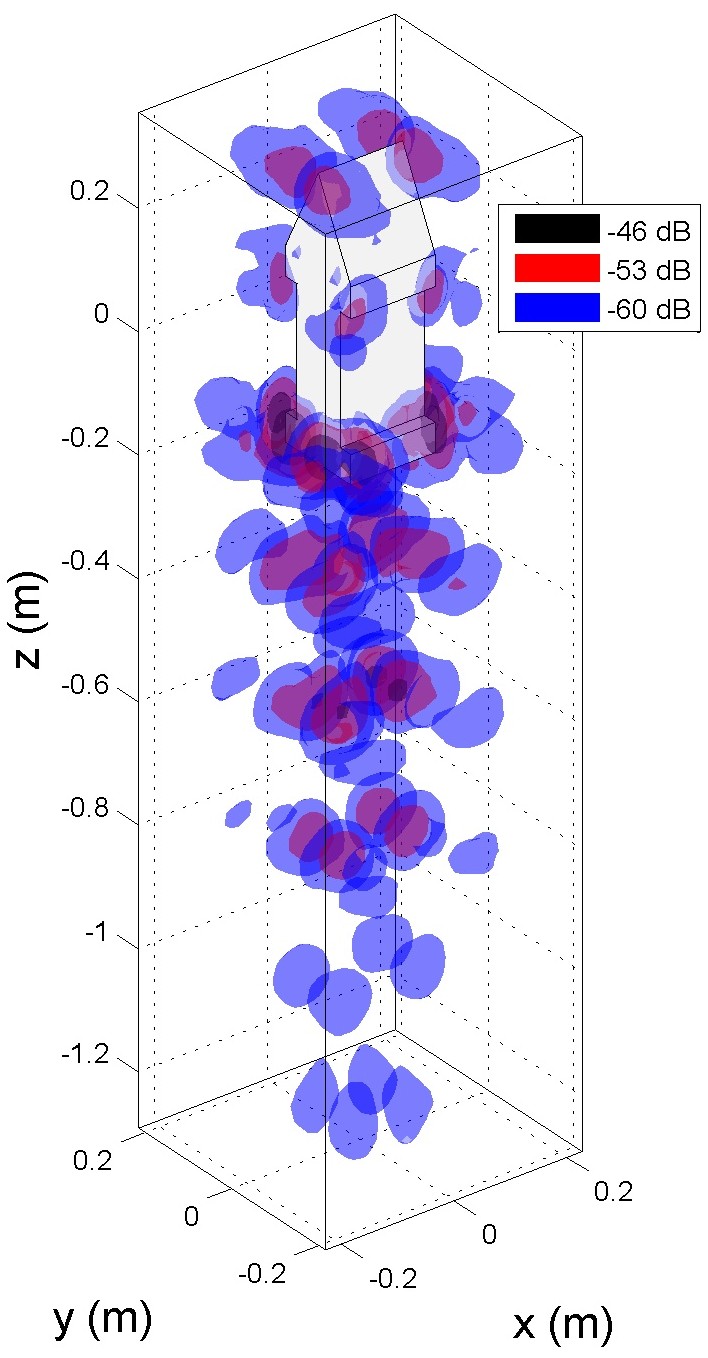}}
\end{center}
\caption{3D radar image of the arrow (from real data): $\wh{\bm{s}}^\textrm{yy}\MCNM$ (left) and $\wh{\bm{s}}^\textrm{xy}\MCNM$ (right) [$N=40\times40\times1146,M\approx 4\cdot10^6$]}\label{fleche3D_xxxy}
\end{figure}

\begin{figure}[!ht]
\centering \includegraphics[width=0.23\textwidth, clip=true]{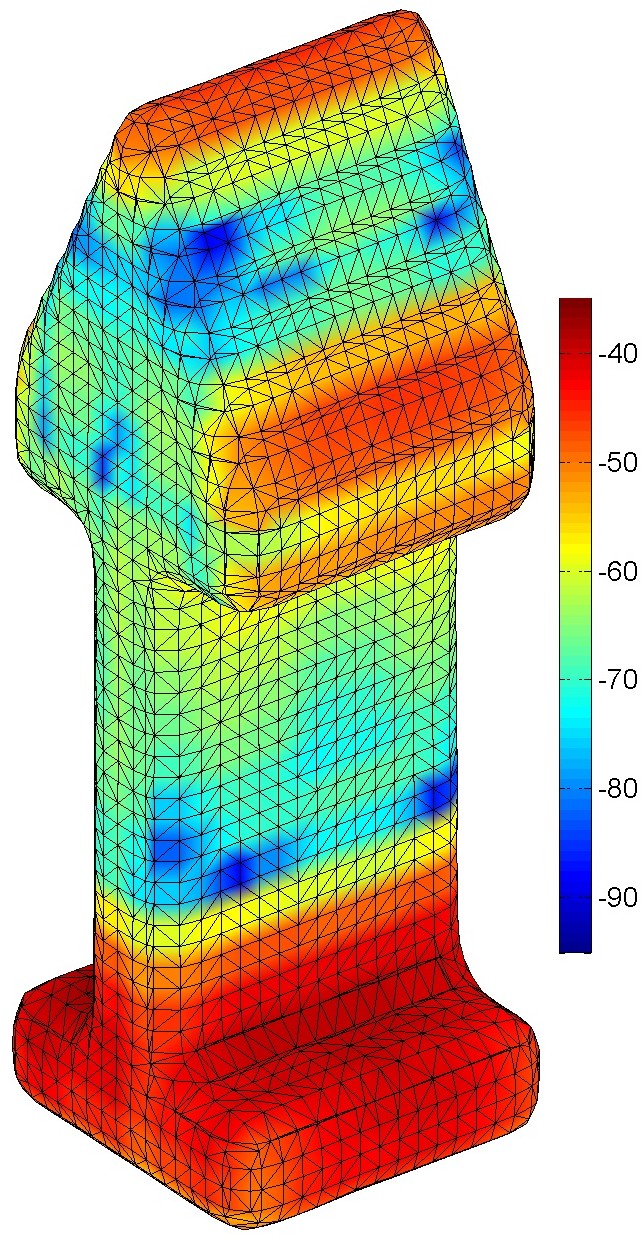} \caption{3D radar image of the arrow:  $\wh{\bm{s}}^\textrm{xx}\MCNM$ (shape section) [$N=40\times40\times1146,M\approx 4\cdot10^6$]} \label{fleche3D}
\end{figure}

Fig.~\ref{fleche3D_yy} provides a comparison of  3D radar maps of the arrow in polarization $\textrm{xx}$. Radar imaging from real data is compared to radar imaging from simulated data, computed with a parallelized 3D EM solver (see \cite{augonnet2014accelerating} for details). It can be checked that both maps are very close one another.  The sensitivity to wave polarization is emphasized in  Fig.~\ref{fleche3D_yy} and \ref{fleche3D_xxxy}. The $\textrm{xx}$ polarization map of  Fig.~\ref{fleche3D_yy}  is again represented in the shape section of Fig. \ref{fleche3D}. Depending on the polarization (i.e. $\textrm{xx}$, $\textrm{yy}$ or $\textrm{xy}$), the diffraction on the edges is more or less important, as well as the multiple reflections of the electromagnetic wave between the rear of the arrow and its base. Located below the target, they are much more important in $\wh{\bm{s}}^\textrm{xx}\MCNM$ and $\wh{\bm{s}}^\textrm{xy}\MCNM$, due to the higher resonance with the base. Let us stress that the 3D Radar imaging $\wh{\bm{s}}^\textrm{xy}\MCNM$  enables to localize not only 3D scatterers associated to diffraction and reflexion but also scatterers that depolarize the incident wave, like the corners of the target. So the simultaneous determination of the three 3D scatterer maps is particularly helpful for RCS analysis: it permits to localize and characterize each 3D backscatterer.


\subsubsection{Near field effects} They can be shown with the full aluminum glider-shaped mock-up of Fig.~\ref{Arch3D}, and its large wingspread. Its main dimensions are approximately 3 ft $\times$ 3 ft $\times$ 8 in (see \cite{Chevalier} for details). The glider mock-up is located vertically on the polystyrene mast. 
 
\begin{figure}[!ht]
\centering \includegraphics[width=0.5\textwidth, clip=true]{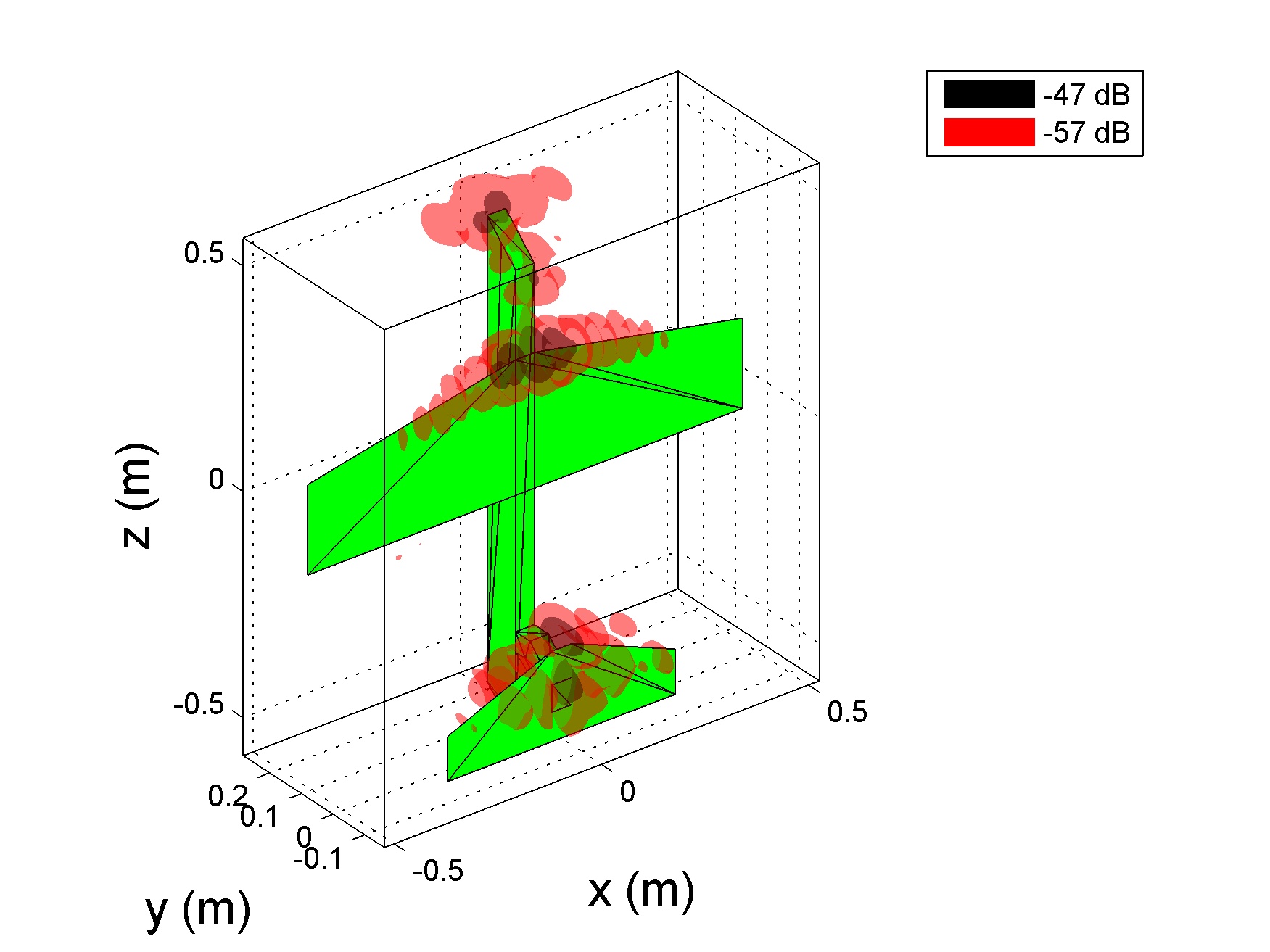} \caption{3D radar image of the glider: $\wh{\bm{s}}^\textrm{xx}\MCNM$ [$N=128\times128\times128,M\approx 1.3\cdot10^5$]} \label{planeur_xx}
\end{figure}

\begin{figure}[!ht]
\centering \includegraphics[width=0.5\textwidth, clip=true]{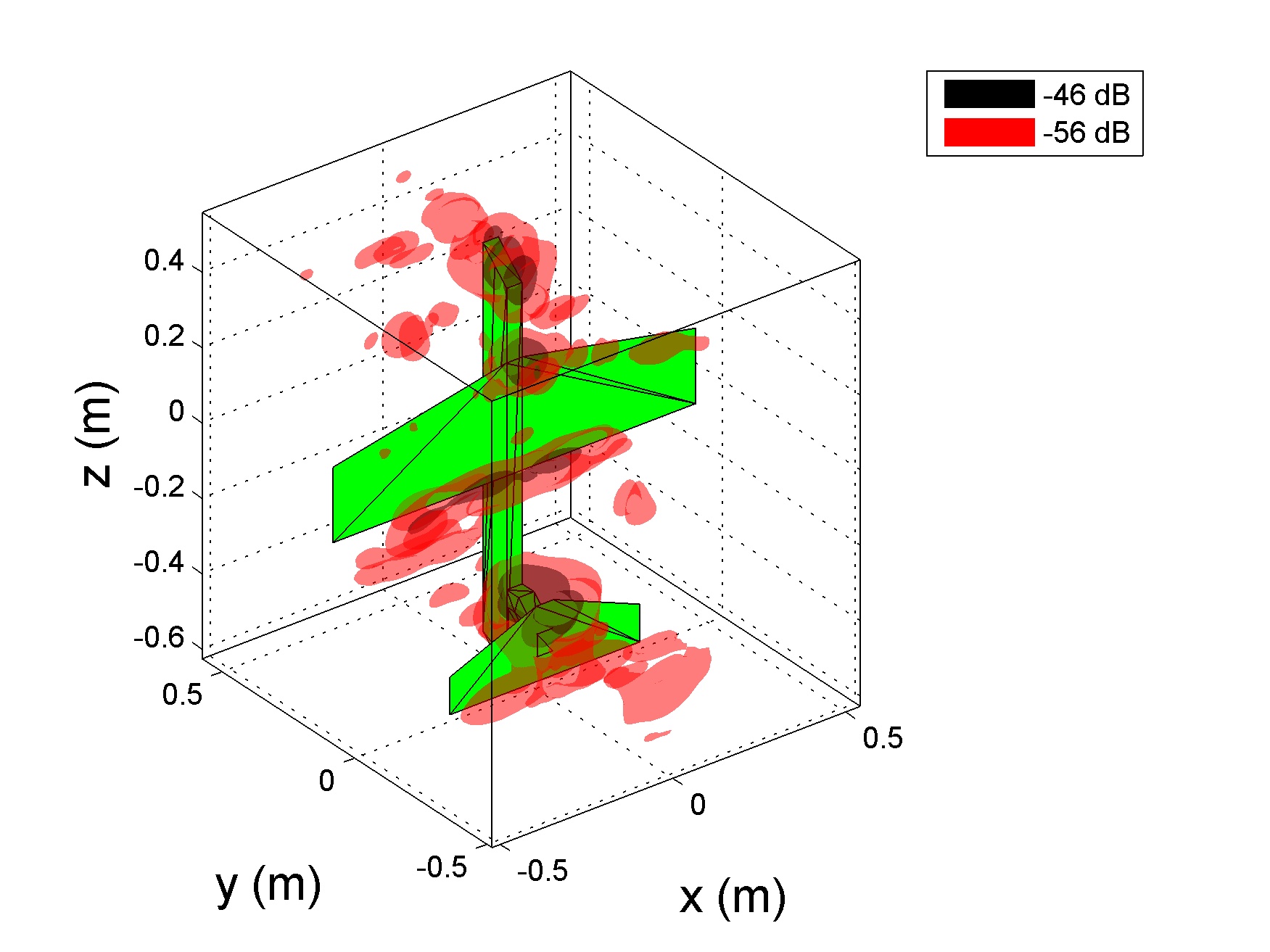} \caption{3D radar image of the glider: $\wh{\bm{s}}^\textrm{yy}\MCNM$ [$N=128\times128\times128,M\approx 1.3\cdot10^5$]} \label{planeur_yy}
\end{figure}

In Fig.~\ref{planeur_xx} and \ref{planeur_yy},  3D scatterer maps are shown depending on polarization, respectively $\textrm{xx}$ and $\textrm{yy}$. Even if the maps are a bit more complex with several multiple scattering artifacts, the main scatterers of the target are apparent: the nose, the fuselage and the tail. However, it can be clearly seen that it does not exhibit the expected high specular behavior of the trailing edge of the front wing glider. Contrary to far field, the backscatterrer is not located uniformly along the entire length of the wing. It  mainly results from aliasing, due to angular undersampling both in $\theta$ and $\varphi$, in the context of the extended wingspread in the transversal dimension. It also partly results  from  the spherical RCS data and the anisotropic radiation pattern of the antenna \cite{Chevalier}. These defaults can be corrected by improving the angular sampling and by introducing a 3D near field/far field correction in the radar imaging algorithm. 

\section{Conclusion} \label{Section_Conclusion}

A 3D radar imaging technique has been presented. It is able to process a collected scattered field data made of polarization-diverse measurements, where the electric field direction varies during the backscatter data acquisition. It is based on  fast and efficient regularized inversion that reconstruct three huge 3-D scatterer maps at a time. The approach rests on a simple but original extension of the standard multiple scatterer point model, closely related to HR polarimetric characterization. It is applied successfully to synthetic and real data, collected from a 3D spherical experimental layout dedicated to accurate RCS characterization. 

To go further, the resolution could be enhanced by introducing non-quadratic approaches (L2-L1) \cite{Giovannelli05,ccetin2001feature} or considering none differentiable criteria. From \cite{Boubertakh06}, constraints could be introduced, e.g. from the known shape, in order to improve the scatter inference and remove ambiguities \cite{mensa1991high}. Besides, similarly  to \cite{broquetas1998spherical,fortuny1999three}, it could be possible to take into account the spherical near field illumination of the target as well as the radiation pattern of the antenna.

\bibliographystyle{IEEEtran}
\bibliography{ref}

\section*{Acknowledgments}
The authors gratefully acknowledge help from B. Stufel and M. Sesques about electromagnetism features and synthetic data.

\appendix

Consider the general quadratic problem in $\eC^N$ with $P$ linear equality constraints:
\beq \label{Eq:PbIOptimMCNM_2}
\wh{\xb} = 
\bca
\displaystyle \argmin_{\xb\in\eC^N} \norm{\xb}^2_\Qb \cr 
\ST~ \cb-\Ab\xb=0
\eca
\eeq
where 
\bit

\item $\Qb$ is a $N\times N$ positive-definite matrix that defines the criterion to be minimized

\item while the constraint is decribed through the vector $\cb\in\eC^P$ and the $P\times N$ matrix $\Ab$. This matrix is assumed to be of full rank. 

\eit
A standard solution to such a problem relies on Lagrange theory: Lagrange multipliers, duality and saddle point, as follows. The Lagrangian of the problem~(\ref{Eq:PbIOptimMCNM_2}) reads:
\beq\label{Eq:Lagrangien_2}
\Lc(\xb,\ub) =  \xb^\dag \Qb \xb  + \ub^\dag (\cb-\Ab\xb)
\eeq
where $\ub\in\eC^P$ is the Lagrange multiplier, also referred to as the dual variable ($\xb\in\eC^N$ is referred to as the primal variable).  

Let minimize $\Lc$ with respect to $\xb$ (for a fixed $\ub$) by setting to zero the gradient of the Lagrangian: 
\beqx
\displaystyle\frac{\partial \Lc}{\partial \xb} (\bar\xb,\ub) =  2\Qb \bar\xb - \Ab^\dag\ub = 0
\eeqx
The solution is directly given by: 
\beq\label{Eq:PrimalOptimiseur}
\bar\xb =  \frac{1}{2} \Qb\pmu \Ab^\dag\ub \,.
\eeq
Note that the Hessian is equal to $2\Qb$ and hence is strictly positive. Consequently, $\bar\xb$ corresponds to a minimum. Then, the introduction of its value in the expression of $\Lc$ given by~(\ref{Eq:Lagrangien_2}) provides the dual function $\bar\Lc$:
\beqx
\bar\Lc(\ub) = \inf_\xb\Lc(\xb,\ub) = \Lc(\bar\xb,\ub)  
		= -\frac{1}{4} \ub^\dag\Ab\Qb\pmu\Ab^\dag\ub + \ub^\dag \cb \,.
\eeqx
Next, let maximize $\bar\Lc(\ub)$ relatively to $\ub$ by setting to zero also the gradient with respect to the dual variable: 
\beqx
0 = \displaystyle\frac{\partial \bar\Lc}{\partial \ub} (\bar\ub) = -\frac{1}{2}\Ab \Qb\pmu \Ab^\dag \bar\ub + \cb 
\eeqx
The solution is directly given by: %
\beq\label{Eq:DualOptimiseur}
\bar\ub =  2 (\Ab\Qb\pmu\Ab^\dag)\pmu \cb \,.
\eeq
Note that the Hessian is equal to $-(\Ab\Qb\Ab^\dag)/2$ and hence is strictly negative; $\bar\ub$ corresponds to a maximum. 
Let insert the expression~(\ref{Eq:DualOptimiseur}) of the dual optimizer into the expression~(\ref{Eq:PrimalOptimiseur}) of the primal optimizer. It leads to: 
\beq
\wh{\xb}  =  \Qb\pmu\Ab^\dag (\Ab\Qb\pmu\Ab^\dag)\pmu \cb\,
\eeq
which is  the Minimum Norm Least Squares (MNLS) solution. When $\Qb$ is the identity matrix, it results in the expression (\ref{Eq:OptimMCNM}):
\beqx
\wh{\xb}  = \Ab^\dag (\Ab\Ab^\dag)\pmu \cb\,.
\eeqx
Note that, if $N=P$, $\Ab$ is invertible and $\wh{\xb}  = \Ab\pmu \cb$.

\end{document}